\documentstyle[12pt]{article}
\newread\epsffilein    % file to \read
\newif\ifepsffileok    % continue looking for the bounding box?
\newif\ifepsfbbfound   % success?
\newif\ifepsfverbose   % report what you're making?
\newdimen\epsfxsize    % horizontal size after scaling
\newdimen\epsfysize    % vertical size after scaling
\newdimen\epsftsize    % horizontal size before scaling
\newdimen\epsfrsize    % vertical size before scaling
\newdimen\epsftmp      % register for arithmetic manipulation
\newdimen\pspoints     % conversion factor
\def\epsfbox#1{\global\def\epsfllx{72}\global\def\epsflly{72}%
   \global\def\epsfurx{540}\global\def\epsfury{720}%
   \def\lbracket{[}\def\testit{#1}\ifx\testit\lbracket
   \let\next=\epsfgetlitbb\else\let\next=\epsfnormal\fi\next{#1}}%
\def\epsfgetlitbb#1#2 #3 #4 #5]#6{\epsfgrab #2 #3 #4 #5 .\\%
   \epsfsetgraph{#6}}%
\def\epsfnormal#1{\epsfgetbb{#1}\epsfsetgraph{#1}}%
\def\epsfgetbb#1{%
\openin\epsffilein=#1
\ifeof\epsffilein\errmessage{I couldn't open #1, will ignore it}\else
   {\epsffileoktrue \chardef\other=12
    \def\do##1{\catcode`##1=\other}\dospecials \catcode`\ =10
    \loop
       \read\epsffilein to \epsffileline
       \ifeof\epsffilein\epsffileokfalse\else
          \expandafter\epsfaux\epsffileline:. \\%
       \fi
   \ifepsffileok\repeat
   \ifepsfbbfound\else
    \ifepsfverbose\message{No bounding box comment in #1; using defaults}\fi\fi
   }\closein\epsffilein\fi}%
\def\epsfsetgraph#1{%
   \epsfrsize=\epsfury\pspoints
   \advance\epsfrsize by-\epsflly\pspoints
   \epsftsize=\epsfurx\pspoints
   \advance\epsftsize by-\epsfllx\pspoints
   \epsfxsize\epsfsize\epsftsize\epsfrsize
   \ifnum\epsfxsize=0 \ifnum\epsfysize=0
      \epsfxsize=\epsftsize \epsfysize=\epsfrsize
     \else\epsftmp=\epsftsize \divide\epsftmp\epsfrsize
       \epsfxsize=\epsfysize \multiply\epsfxsize\epsftmp
       \multiply\epsftmp\epsfrsize \advance\epsftsize-\epsftmp
       \epsftmp=\epsfysize
       \loop \advance\epsftsize\epsftsize \divide\epsftmp 2
       \ifnum\epsftmp>0
          \ifnum\epsftsize<\epsfrsize\else
             \advance\epsftsize-\epsfrsize \advance\epsfxsize\epsftmp \fi
       \repeat
     \fi
   \else\epsftmp=\epsfrsize \divide\epsftmp\epsftsize
     \epsfysize=\epsfxsize \multiply\epsfysize\epsftmp
     \multiply\epsftmp\epsftsize \advance\epsfrsize-\epsftmp
     \epsftmp=\epsfxsize
     \loop \advance\epsfrsize\epsfrsize \divide\epsftmp 2
     \ifnum\epsftmp>0
        \ifnum\epsfrsize<\epsftsize\else
           \advance\epsfrsize-\epsftsize \advance\epsfysize\epsftmp \fi
     \repeat
   \fi
   \ifepsfverbose\message{#1: width=\the\epsfxsize, height=\the\epsfysize}\fi
   \epsftmp=10\epsfxsize \divide\epsftmp\pspoints
   \vbox to\epsfysize{\vfil\hbox to\epsfxsize{%
      \includegraphics{#1}%
      \hfil}}%
\epsfxsize=0pt\epsfysize=0pt}%

{\catcode`\%=12 \global\let\epsfpercent=%\global\def\epsfbblit{%BoundingBox}}%
\long\def\epsfaux#1#2:#3\\{\ifx#1\epsfpercent
   \def\testit{#2}\ifx\testit\epsfbblit
      \epsfgrab #3 . . . \\%
      \epsffileokfalse
      \global\epsfbbfoundtrue
   \fi\else\ifx#1\par\else\epsffileokfalse\fi\fi}%
\def\epsfgrab #1 #2 #3 #4 #5\\{%
   \global\def\epsfllx{#1}\ifx\epsfllx\empty
      \epsfgrab #2 #3 #4 #5 .\\\else
   \global\def\epsflly{#2}%
   \global\def\epsfurx{#3}\global\def\epsfury{#4}\fi}%
\def\epsfsize#1#2{\epsfxsize}
\let\epsffile=\epsfbox

\def\ltap{\raisebox{-.4ex}{\rlap{$\sim$}} \raisebox{.4ex}{$<$}}
\def\gtap{\raisebox{-.4ex}{\rlap{$\sim$}} \raisebox{.4ex}{$>$}}

\newlength{\dinwidth}
\newlength{\dinmargin}
\setlength{\dinwidth}{21.0cm}
\textheight24.2cm \textwidth17.0cm
\setlength{\dinmargin}{\dinwidth}
\addtolength{\dinmargin}{-\textwidth}
\setlength{\dinmargin}{0.5\dinmargin}
\oddsidemargin -1.0in
\addtolength{\oddsidemargin}{\dinmargin}

\setlength{\evensidemargin}{\oddsidemargin}
\setlength{\marginparwidth}{0.9\dinmargin}
\marginparsep 8pt \marginparpush 5pt
\topmargin -42pt
\headheight 12pt
\headsep 30pt %\footheight 12pt \footskip 24pt
\parskip 3mm plus 2mm minus 2mm
\parindent 5mm

\newcommand{\bfig}{\begin{figure}}
\newcommand{\efig}{\end{figure}}
\newcommand{\bcen}{\begin{center}}
\newcommand{\ecen}{\end{center}}
\newcommand{\beq}{\begin{equation}}
\newcommand{\eeq}{\end{equation}}
\newcommand{\btabu}{\begin{tabular}}
\newcommand{\etabu}{\end{tabular}}
\newcommand{\btabl}{\begin{table}}
\newcommand{\etabl}{\end{table}}

\newcommand{\QSQ}{\mbox{$Q^2$}}

\def\lsim{\mathrel{\rlap{\lower4pt\hbox{\hskip1pt$\sim$}}
    \raise1pt\hbox{$<$}}}         %less than or approx. symbol
\def\gsim{\mathrel{\rlap{\lower4pt\hbox{\hskip1pt$\sim$}}
    \raise1pt\hbox{$>$}}}         %greater than or approx. symbol

\begin{document}

\title { \vspace{2cm}
{\bf  Measurement of Elastic $\rho^0$
 Photoproduction at HERA }
\\
\author{\rm The ZEUS Collaboration \\}
}
\date{ }
\maketitle

\vspace{5 cm}
\begin{abstract}
Elastic $\rho^0$ photoproduction has been measured using the ZEUS detector
at HERA. Untagged photoproduction events from $ep$ interactions were used
to measure the reaction $\gamma p \rightarrow \rho^0 p$ ($\rho^0
\rightarrow \pi^+ \pi^-$) at photon-proton centre-of-mass energies between
60 and 80~GeV and $|t|<0.5$~GeV$^2$, where $t$ is the square of the
four-momentum transferred at the proton vertex. The differential cross
section $d\sigma/dM_{\pi\pi}$, where $M_{\pi\pi}$ is the invariant mass of
the two pions, and the integrated cross section, $\sigma_{\gamma
p\rightarrow \rho^0 p}$, are presented; the latter was measured to be
$14.7\pm 0.4~(\mbox{stat.})~\pm2.4~(\mbox{syst.})~\mu\mbox{b}$.  The
differential cross section $d\sigma/dt$ has an approximately exponential
shape; a fit of the type $A^{\prime}_{t} \exp{(-b^{\prime}_{t}|t| +
c^{\prime}_{t} t^2)}$ yields a $t$-slope $b^{\prime}_{t}=
9.9~\pm~1.2~(\mbox{stat.})~\pm 1.4~(\mbox{syst.})~\mu\mbox{b}$.  The
results, when compared to low energy data, show a weak energy dependence
of both $\sigma_{\gamma p\rightarrow \rho^0 p}$ and of the $t$-slope. The
$\rho^0$ is produced predominantly with transverse polarisation,
demonstrating that $s$-channel helicity conservation holds at these
energies.
\end{abstract}

\vspace{-20cm}
\begin{flushleft}
\tt DESY 95-143 \\
July 1995 \\
\end{flushleft}

\setcounter{page}{0}
\thispagestyle{empty}
\newpage

%   19/07/95            MEMBER NAME  AUTH27   (TEX)      M  TEX
%   11/07/95 507151928  MEMBER NAME  AUTH027  (ZEUS)     M  TEX
%
\def\3{\ss}
\parindent 0cm
\footnotesize
\renewcommand{\thepage}{\Roman{page}}
\begin{center}
\begin{large}
The ZEUS Collaboration
\end{large}
\end{center}
M.~Derrick, D.~Krakauer, S.~Magill, D.~Mikunas, B.~Musgrave,
J.~Repond, R.~Stanek, R.L.~Talaga, H.~Zhang \\
{\it Argonne National Laboratory, Argonne, IL, USA}~$^{p}$\\[6pt]
R.~Ayad$^1$, G.~Bari, M.~Basile,
L.~Bellagamba, D.~Boscherini, A.~Bruni, G.~Bruni, P.~Bruni, G.~Cara
Romeo, G.~Castellini$^{2}$, M.~Chiarini,
L.~Cifarelli$^{3}$, F.~Cindolo, A.~Contin, M.~Corradi,
I.~Gialas$^{4}$,
P.~Giusti, G.~Iacobucci, G.~Laurenti, G.~Levi, A.~Margotti,
T.~Massam, R.~Nania, C.~Nemoz, \\
F.~Palmonari, A.~Polini, G.~Sartorelli, R.~Timellini, Y.~Zamora
Garcia$^{1}$,
A.~Zichichi \\
{\it University and INFN Bologna, Bologna, Italy}~$^{f}$
% ...
\\[6pt]
\clearpage
\hspace*{1mm}
$^{ 1}$ supported by Worldlab, Lausanne, Switzerland \\
\hspace*{1mm}
$^{ 2}$ also at IROE Florence, Italy  \\
\hspace*{1mm}
$^{ 3}$ now at Univ. of Salerno and INFN Napoli, Italy  \\
\hspace*{1mm}
$^{ 4}$ supported by EU HCM contract ERB-CHRX-CT93-0376 \\
\hspace*{1mm}
% ...
$^{18}$ now with OPAL Collaboration, Faculty of Physics at Univ. of
        Freiburg \\
% ...

\begin{tabular}{lp{15cm}}
$^{a}$ &supported by the Natural Sciences and Engineering Research
         Council of Canada (NSERC) \\
$^{b}$ &supported by the FCAR of Qu\'ebec, Canada\\
$^{c}$ &supported by the German Federal Ministry for Research and
         Technology (BMFT)\\
$^{d}$ &supported by the MINERVA Gesellschaft f\"ur Forschung GmbH,
         and by the Israel Academy of Science \\
$^{e}$ &supported by the German Israeli Foundation, and
         by the Israel Academy of Science \\
$^{f}$ &supported by the Italian National Institute for Nuclear Physics
         (INFN) \\
$^{g}$ &supported by the Japanese Ministry of Education, Science and
         Culture (the Monbusho)
         and its grants for Scientific Research\\
$^{h}$ &supported by the Korean Ministry of Education and Korea Science
         and Engineering Foundation \\
$^{i}$ &supported by the Netherlands Foundation for Research on Matter
         (FOM)\\
$^{j}$ &supported by the Polish State Committee for Scientific Research
         (grant No. SPB/P3/202/93) and the Foundation for Polish-
         German Collaboration (proj. No. 506/92) \\
$^{k}$ &supported by the Polish State Committee for Scientific
         Research (grant No. PB 861/2/91 and No. 2 2372 9102,
         grant No. PB 2 2376 9102 and No. PB 2 0092 9101) \\
$^{l}$ &partially supported by the German Federal Ministry for
         Research and Technology (BMFT) \\
$^{m}$ &supported by the German Federal Ministry for Research and
         Technology (BMFT), the Volkswagen Foundation, and the Deutsche
         Forschungsgemeinschaft \\
$^{n}$ &supported by the Spanish Ministry of Education and Science
         through funds provided by CICYT \\
$^{o}$ &supported by the Particle Physics and Astronomy Research
        Council \\
$^{p}$ &supported by the US Department of Energy \\
$^{q}$ &supported by the US National Science Foundation
\end{tabular}

\newpage
\pagenumbering{arabic}
\setcounter{page}{1}
\normalsize

\section{Introduction}

``Elastic" or ``exclusive" production of $\rho^0$ mesons by photons,
$\gamma p \rightarrow \rho^0 p$, has been extensively studied in fixed
target experiments up to photon-proton centre-of-mass energies $W\simeq$
20~GeV, using both real and virtual photons~\cite{bible}-\cite{nmc}.
Recently, the cross section for this reaction has also been obtained in an
indirect measurement at the HERA $ep$ collider, using quasi-real photons
with space-like virtuality $Q^2$ between $4\cdot 10^{-8}$ and $2\cdot
10^{-2}$ GeV$^2$, at an average centre-of-mass energy $\langle W \rangle$
of 180 GeV~\cite{maciek}.

At $W$ values up to about 20 GeV, elastic photoproduction of $\rho^0$
mesons has the characteristic features of soft diffractive processes:  the
dependence of the production cross section on $W$ is weak, the dependence
on $t$, the square of the four-momentum transferred at the proton vertex,
is approximately exponential, and the vector meson is observed to retain
the helicity of the photon ($s$-channel helicity conservation, SCHC).
Such energy and $t$ dependences are also characteristic of hadronic
diffractive processes. The similarity between $\rho^0$ photoproduction and
hadronic processes can be understood in the framework of the Vector
Dominance Model (VDM)~\cite{saku}, in which the photon is assumed to
fluctuate into a vector meson before interacting with the target nucleon;
the reaction $\gamma p \rightarrow \rho^0 p $ is thus related to the
elastic process $\rho^0 p \rightarrow \rho^0 p $.

At sufficiently high energies, diffractive interactions are usually
described in terms of the exchange of a pomeron, an object with the
quantum numbers of the vacuum.  Regge theory provides a framework in which
many of the features of hadronic reactions can be
described~\cite{goulianos}.  In particular, the energy dependence of
diffractive processes is related to the intercept of the pomeron
trajectory.  Several models offer a description of diffractive vector
meson production~\cite{dl}-\cite{ginzburg}; some of them are in the
framework of perturbative QCD. The study of vector meson photoproduction
at the energies available at HERA may thus help to clarify the nature of
the pomeron.

This paper reports a measurement of the elastic $\rho^0$ photoproduction
cross section at $\langle W \rangle$ of 70 GeV, based on about 6,000 $ep
\rightarrow ep \pi^+ \pi^-$ events collected by the ZEUS experiment in
1993. In this measurement the final state electron and proton are not
detected and the relevant kinematic quantities are determined from the
measured three-momenta of the $\rho^0$ decay products, assuming that they
are pions.

The paper is organised as follows. After defining the variables relevant
to $\rho^0$ production, we describe the experimental conditions and the
event selection criteria, and then discuss the acceptance corrections and
the background. From the analysis of the differential cross section
$d\sigma/dM_{\pi\pi}$, where $M_{\pi\pi}$ is the invariant mass of the
$\pi^+ \pi^-$ system, we obtain the integrated cross section
$\sigma_{\gamma p\rightarrow \rho^0 p}$. We then discuss the differential
cross section $d\sigma/dt$ and the angular distributions of the decay
pions. Finally, from the value of $d\sigma/dt$ at $t=0$, the total $\rho^0
p$ cross section is derived using the optical theorem and assuming VDM.

\section{Elastic $\rho^0$ photoproduction at HERA}

\begin{figure}[t]
\begin{center}
\leavevmode
\hbox{%
\epsfxsize =12cm
\epsffile{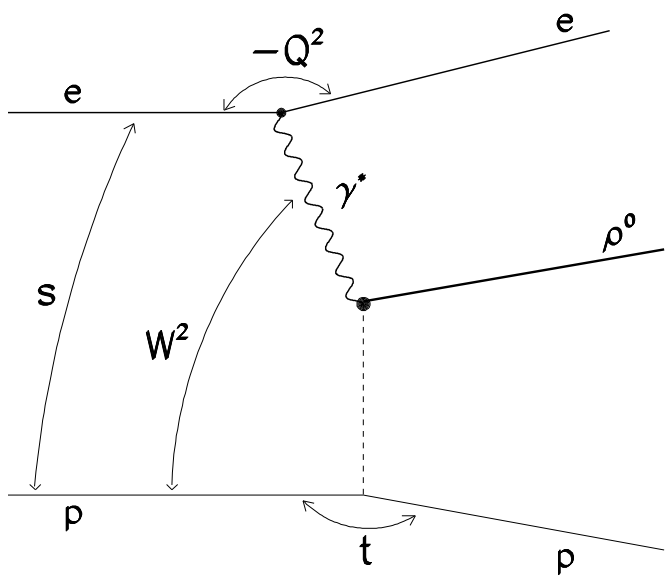}}
\end{center}
\vspace{-4.7cm}
\caption{Elastic $\rho^0$ production in $ep$ collision; the exchanged
virtual photon is indicated
as $\gamma^*$.}
\label{rhodiag}
\end{figure}

Elastic $\rho^0$ photoproduction was investigated by means of the reaction
(see Fig. \ref{rhodiag}) $$ e(k)~ + p(P) \rightarrow e(k') + \rho^0(V) + p
(P'), $$ where the symbols in parenthesis indicate the four-momenta of the
particles involved.

For unpolarised electrons and protons,
two independent variables describe inclusive $ep$ scattering,
since the $ep$ centre-of-mass energy $\sqrt s
=2\sqrt{E_eE_p} = 296$ GeV
is fixed by the energies of the electron ($E_e$) and of the proton ($E_p$)
beams. The variables can be any two of the following four:
\begin{itemize}
\item
 $-Q^2 = q^2 = (k-k')^2$, the four-momentum squared carried by the virtual
 photon;
\item
 $x =Q^2/(2P\cdot q)$, the Bjorken variable;
\item
$y = (q\cdot P)/(k\cdot P)~,$
the fraction of the electron energy transferred by the photon to the
hadronic final state, measured in the proton rest frame;
\item
$W$, the centre-of-mass energy of the $\gamma^* p$ system,
where $$W^2 = (q+P)^2 = -Q^2 + 2y(k\cdot P)+M^2_p,$$
$M_p$ being the proton mass.
\end{itemize}

The hadronic final state, containing the scattered proton and the pions
from the decay $\rho^0 \rightarrow \pi^+ \pi^-$, is described by
additional variables, including the invariant mass $M_{\pi\pi}$ of the two
decay pions, the square of the four-momentum transferred at the proton
vertex, $t$, and the polar and azimuthal angles, defined in
section~\ref{angular}, of the decay pions in the $\pi \pi$ centre-of-mass
frame.

For the data presented here, only the three-momenta of the final state
pions were measured. Events in which the scattered electron was detected
in the ZEUS calorimeter were rejected, thereby restricting $Q^2$ to be
below $Q^2_{\max}\approx 4$~GeV$^2$. The median $Q^2$ is approximately
$10^{-4}$~GeV$^2$.  To explain how the relevant kinematic quantities are
obtained from the four-momenta of the two pions, we first consider the
case $Q^2=Q^2_{\min}= M_e^2 \frac{y^2}{1-y}$ ($M_e$ is the electron mass)
and then discuss the effect of larger $Q^2$. For $Q^2=Q^2_{\min}$
($\approx 10^{-9}$~GeV$^2$ for the kinematic range covered by the present
data), the virtual photon is emitted with zero transverse momentum and
with longitudinal momentum $p_{Z \gamma}\simeq - E_\gamma$ in the
direction opposite to that of the proton beam\footnote{Throughout this
paper we use the standard ZEUS right-handed coordinate system, in which $X
= Y = Z = 0$ is the nominal interaction point, the positive $Z$-axis
points in the direction of flight of the protons (referred to as the
forward direction) and the $X$-axis is horizontal, pointing towards the
centre of HERA.}. Energy and momentum conservation relate the photon
energy $E_{\gamma}$ to the two-pion system energy $E_{\pi \pi}$ and
longitudinal momentum $p_{Z\pi \pi}$ by $2E_{\gamma} \simeq (E_{\pi \pi} -
p_{Z\pi \pi})$. The photon-proton centre-of-mass energy can then be
expressed as:

$$ W^2 \simeq 4 E_\gamma E_p   \simeq
2 (E_{\pi \pi} - p_{Z\pi \pi}) E_p.$$

\noindent
The $\rho^0$ transverse momentum squared in the
laboratory frame, $p_T^2$, approximates to $-t$:
\begin{eqnarray}
   t & = & (q-V)^2 = -\QSQ -2q\cdot V + M^2_{\pi \pi}          \nonumber \\
     & \simeq & -2E_\gamma(E_{\pi \pi}+ p_{Z \pi \pi}) + M^2_{\pi \pi}
\nonumber \\
     & \simeq & -(E^2_{\pi \pi} - p^2_{Z \pi \pi}) + M^2_{\pi \pi}
\nonumber \\
     & = & -p^2_T,                                             \nonumber
\label{tvspt}
\end{eqnarray}
\noindent
where, in addition to $2E_{\gamma} \simeq  (E_{\pi \pi} - p_{Z\pi \pi})$,
we have used the approximation  $Q^2=0$.
Non-zero values of $Q^2$ cause $p_T^2$ to differ from $-t$
by $\ltap Q^2$; the effect on the measured distributions is
discussed in section~\ref{ptt}.
For this measurement, the minimum kinematically allowed
value of $|t|$ is negligible, $|t_{\min}|\simeq 10^{-8}$~GeV$^2$ at $W=70$~GeV.

Fig.~\ref{papercomp} shows the scatter
plot of the reconstructed versus the true values of $W$ and $t$
for the sample of Monte Carlo
events used to evaluate the acceptance (cf. section~\ref{ptt}).
The difference of the reconstructed and the true value of $W$ has a mean value
of
zero and an r.m.s. spread
of 1.4~GeV. The analogous difference for $t$ has also a mean value of
approximately zero
and an r.m.s. spread of 0.06~GeV$^2$. The events far from the
diagonal have $Q^2 \gg Q^2_{\min}$.
Throughout  the analysis, the variable $W$ was calculated
using the above approximation. For $t$, the $Q^2$ dependence of the
relation between  $t$ and $p_T^2$ was taken into account in the acceptance
correction, as discussed in section~\ref{ptt}.

In the one photon exchange approximation,
the $e p$ and the $\gamma^* p$ cross sections for elastic $\rho^0$
production are related by
\begin{eqnarray}
 \frac{d^2\sigma_{ep \rightarrow ep\rho^0}}{dydQ^2} = \frac{\alpha}{2\pi Q^2}
   \left[\left( \frac{1+(1-y)^2}{y} - \frac{2(1-y)}{y}
   \cdot \frac{Q_{\min}^2}{Q^2}\right) \cdot
   \sigma_T^{\gamma^*p \rightarrow \rho^0 p}(W,Q^2) + \right. \nonumber \\
   \left. \frac{2(1-y)}{y} \cdot
   \sigma_L^{\gamma^*p \rightarrow \rho^0 p}(W,Q^2)\right],
\label{bornc}
\end{eqnarray}
where $\alpha$ is the fine structure constant and
$\sigma_T^{\gamma^*p \rightarrow \rho^0 p}(W,Q^2)$ and
$\sigma_L^{\gamma^*p\rightarrow \rho^0 p}(W,Q^2)$
are the respective cross sections for transversely and longitudinally
polarised virtual photons.
Following VDM, these are related by
\begin{eqnarray}
\sigma_L^{\gamma^*p \rightarrow \rho^0 p}(W,Q^2) \simeq
\sigma_T^{\gamma^*p \rightarrow \rho^0 p}(W,Q^2)
\cdot \frac{Q^2}{M_{\rho}^2},
\label{sigmal}
\end{eqnarray}

\noindent
with
\begin{eqnarray}
\sigma_T^{\gamma^*p \rightarrow \rho^0 p}(W,Q^2) =
\sigma_{\gamma p\rightarrow \rho^0 p}(W) \left/
\left(1+\frac{Q^2}{M_{\rho}^2}\right)^2 \right.,
\label{sigmat}
\end{eqnarray}

\noindent
where $\sigma_{\gamma p \rightarrow \rho^0 p}(W)$ is the
cross section for elastic photoproduction ($Q^2=0$) of $\rho^0$ mesons,
and $M_{\rho}$ is the $\rho^0$ meson mass.
Substituting the latter two expressions into equation~(\ref{bornc})
yields:
\begin{eqnarray}
\frac{d^2\sigma_{ep \rightarrow ep \rho^0}}{dy dQ^2}
=\Phi(y,Q^2) \cdot \sigma_{\gamma p \rightarrow \rho^0 p }(W(y)),
\label{crs}
\end{eqnarray}

\noindent
where the function
\begin{eqnarray}
\Phi(y,Q^2)=\frac{\alpha}{2 \pi Q^2}
\left\{\left[ \frac{1+(1-y)^2}{y} - \frac{2(1-y)}{y}
\left(\frac{Q_{\min}^2}{Q^2} - \frac{Q^2}{M_{\rho}^2} \right) \right]
\frac{1 }{\left(1+\frac{Q^2}{M_{\rho}^2}\right)^2}
\right\}
\label{flux}
\end{eqnarray}
\noindent
is the effective photon flux.

The cross section $\sigma_{\gamma p \rightarrow \rho^0 p}(\langle W
\rangle)$ for elastic $\rho^0$ photoproduction is thus obtained as the
ratio of the corresponding acceptance corrected electron-proton
cross-section, integrated over the $y$ and $Q^2$ ranges covered by the
measurement, and the effective photon flux $\Phi(y,Q^2)$ integrated over
the same $y$ and $Q^2$ ranges.  This procedure determines the cross
section, assuming VDM, for elastic production of $\rho^0$ mesons at
$Q^2=0$ averaged over the $W$ range of the measurement.

\section{Experimental conditions}

\subsection{HERA}

During 1993, HERA operated at a proton energy of $820$ GeV
and an electron energy of $26.7$~GeV; 84 colliding
electron-proton bunches were stored, with an additional 6 unpaired
proton and 10 unpaired electron bunches. These additional
pilot bunches were used for background studies.
The time between bunch crossings was $96$~ns. Typical bunch currents
were 10~mA for both the electron and the proton beam, providing
luminosities of approximately $6\cdot10^{29}$~cm$^{-2}$~s$^{-1}$.

\subsection{The ZEUS detector}

The components of the ZEUS detector are described in detail in
ref.~\cite{status93}. A short description of those most relevant to the
present analysis follows.

Charged particles are tracked by the vertex detector (VXD) and the central
tracking detector (CTD) which operate in a magnetic field of 1.43 T
provided by a thin superconducting solenoid.  The VXD \cite{VXD} is a
cylindrical drift chamber that surrounds the beam pipe and consists of 120
radial cells, each with 12 sense wires. The VXD resolution in the plane
transverse to the beam direction, for the data presented here, is 50
$\mu$m in the central region of a cell and 150 $\mu$m near the cell edges.
The CTD~\cite{CTD} consists of 72 cylindrical drift chamber layers,
organised in 9 superlayers covering the polar angle region $15^\circ <
\theta < 164^\circ$. In 1993, the spatial resolution in the plane
perpendicular to the beam was $\simeq 260~\mu$m. For the data presented in
this paper, the combined CTD and VXD information provides resolutions for
the primary $ep$ interaction vertex of 1.1 cm in $Z$ and 0.2 cm in the
$XY$ plane. The momentum resolution, for tracks traversing all
superlayers, is $\sigma(p_t)/p_t \approx \sqrt{(0.005)^2p_t^2 +
(0.016)^2}$, where $p_t$ is in GeV.

The high resolution uranium-scintillator calorimeter (CAL) \cite{CAL}
consists of a forward (FCAL), a barrel (BCAL) and a rear (RCAL) part,
respectively covering the polar regions $2.2^\circ$ to $36.7^\circ$,
$36.7^\circ$ to $129.1^\circ$, and $129.1^\circ$ to $176.6^\circ$. The
calorimeter parts are subdivided transversely into towers and
longitudinally into electromagnetic (EMC) and hadronic (HAC) sections. A
section of a tower is called a cell; each cell is viewed by two
photomultiplier tubes. Holes of $20 \times 20$~cm$^2$ in the centre of
FCAL and RCAL accommodate the HERA beam pipe. From test beam data, the
energy resolution was found to be $ \sigma_E/E = 0.18/\sqrt{E(\mbox{GeV})}
$ for electrons and $\sigma_E/E = 0.35/\sqrt{E(\mbox{GeV})}$ for hadrons.
The performance, energy and time calibration of the calorimeter are
continuously monitored using pedestal triggers, charge and light injection
as well as the uranium radioactivity.  The additional features relevant to
the present analysis are the sub-nanosecond time resolution and the low
noise of approximately 15 MeV for the electromagnetic and 25 MeV for the
hadronic cells.

The veto wall is used to tag events in which a proton has scattered off a
residual gas molecule in the beam pipe (``proton-gas" events) upstream of
the ZEUS detector. It consists of two layers of scintillators, with
overall dimensions of 500~cm $\times$ 600~cm, on both sides of an 87~cm
thick iron wall centred at $Z= -7.3$~m.

The C5 beam monitor, a small lead-scintillator counter assembly located at
$Z=-3.2$~m, records the arrival times of halo particles associated with
the proton and electron bunches within $10$ cm of the beam axis. It is
used to verify the relative timing of the beams and to reject events due
to proton-gas interactions.

The luminosity detector (LUMI) \cite{lumi} measures the luminosity by
means of the Bethe-Heitler reaction $ep\rightarrow ep \gamma$; a detailed
description of the method used is given in~\cite{maciek}. The
bremsstrahlung events are identified by measuring the radiated photon in a
lead-scintillator calorimeter placed in the HERA tunnel downstream of the
interaction point in the direction of the outgoing electron.

\subsection{Untagged $\rho^0$ photoproduction trigger}

ZEUS uses a three level trigger system \cite{status93}. The data presented
here come from the ``untagged photoproduction trigger'', designed to
select vector meson photoproduction events \cite{tesi}. The term
``untagged'' refers to the fact that the scattered electron escapes
undetected through the beam pipe hole in the RCAL and is not detected in
the LUMI detector.

The trigger conditions can be summarised as follows:
\begin{itemize}
\item
First-level-trigger (FLT):
\begin{itemize}
\item
At least 464 MeV deposited in the electromagnetic section of RCAL.
\item
At least one track candidate in the CTD.
\item
Less than 3750 MeV deposited in the calorimeter towers surrounding
the beam pipe in the forward direction. This requirement suppressed proton
beam-gas events and a significant fraction of the photoproduction
cross section.
\end{itemize}
The trigger was vetoed if hits were present in the C5 or in the veto wall
counters, with timing consistent with that of a
$p$-gas collision occurred upstream of the interaction point.

The resulting FLT rate was $\approx 10$~Hz
at a luminosity of $6\cdot10^{29}$~cm$^{-2}$~s$^{-1}$.

\item
Second-level-trigger (SLT):
\begin{itemize}
\item
Events with calorimeter timing indicating that the interaction had
occurred upstream of the interaction point were rejected.
\end{itemize}

\item
Third-level-trigger (TLT):
\begin{itemize}
\item
Cosmic ray events were discarded on the basis of calorimeter timing.
\item
A tighter calorimeter timing rejection was applied.
\item
A cut on the $Z$ value of the reconstructed
event vertex of $\pm~85$ cm was imposed.
\end{itemize}
The rate of events passing the untagged photoproduction trigger at the third
level was about 1.2 Hz.
For some fraction of the data-taking period a factor of 3 prescale was applied.
\end{itemize}

The requirements that a signal be detected in RCAL and a track be seen in
the CTD effectively selected events with photon energies between 0.5 and 4
GeV, corresponding to $40 < W < 120$ GeV.

\section{Event selection}
\label{cuts}

During 1993, approximately $7\cdot 10^6$ events were recorded,
corresponding to a total integrated luminosity of about 550 nb$^{-1}$. The
data presented in this paper correspond to an effective luminosity,
accounting for the effects of the prescaling of the trigger mentioned in
the previous section, of $(240 \pm 8)$~nb$^{-1}$.

The following offline requirements were imposed in order to obtain
the final $\rho^0$ photoproduction sample:
\begin{itemize}
\item exactly two tracks from particles of opposite charge and both
associated with a reconstructed vertex;
\item transverse momentum greater than 200 MeV and
hits in at least the 3 innermost CTD superlayers for each of the
two tracks, thus restricting the
data to a region of well understood track reconstruction efficiency;
this restriction approximately translates into one on the track
pseudorapidity ($\eta=-\ln{\tan{\theta/2}}$) of $|\eta|~\ltap~1.8$;
\item vertex position within $\pm$~40~cm of the nominal interaction
point and within a radial distance of 1.5 cm from the beam axis (the
interaction region was centred at $Z=-6$~cm and had an r.m.s.
width of $\pm 11$~cm);
\item total energy in the forward calorimeter $E_{FCAL} \leq$ 1 GeV, thereby
limiting the contamination from the reaction
$e p \rightarrow e   \rho^0 X$, where $X$ is a hadronic state of
mass $M_X$ into which the proton had dissociated;
\item no energy deposits larger than 200 MeV in any calorimeter cell outside
a circular region around the track impact point with a radius of 30~cm
in the EMC and 50~cm in the HAC, thus rejecting events with
neutral particles or with charged particles outside the region of
sensitivity of the tracking system.
\end{itemize}
\noindent
A total of 13570 events satisfied the above criteria.

The pion mass was assigned to each track and the analysis was restricted to
events reconstructed in the kinematic region defined by:
\begin{eqnarray}
               60    < & W & < 80   ~\mbox{GeV},     \nonumber\\
                0.55 < & M_{\pi\pi}  & <  1.0 ~\mbox{GeV},\label{kin}   \\
                       & p_T^2        & <  0.5 ~\mbox{GeV}^2 . \nonumber
\end{eqnarray}
In the chosen energy range the acceptance is well understood.
The restricted mass range reduces the contamination from reactions
involving other vector mesons, in particular from elastic $\phi$
and $\omega$ production, as well as from photon conversions. The restricted
$p_T$ range reduces the contamination from events with
diffractive dissociation of the proton ($ep \rightarrow e\rho^0 X$). The final
sample contains 6381 events.

The invariant mass spectrum is shown in Fig.~\ref{massbef} before and
after the offline selection.  The data are dominated by the $\rho^0$
signal. The corresponding mass spectra of like-sign two track events are
also shown as the shaded areas. The small peak just above the $\pi \pi$
threshold in Figs.~\ref{massbef}b and c is due to $\phi \rightarrow K^+
K^-$ events, where the pion mass has been erroneously assigned to the
tracks.

\section{Monte Carlo generators and acceptance calculation}
\label{ptt}

The reaction $ep \rightarrow e\rho^0p$ was modelled using two different
Monte Carlo ge\-ne\-ra\-tors.

The first generator, DIPSI~\cite{dipsi}, describes $\rho^0$
photoproduction in terms of pomeron exchange.  Based on the model of
Ryskin~\cite{misha}, it assumes that the exchanged photon fluctuates into
a $q\bar{q}$ pair which then interacts with the pomeron emitted by the
incident proton. The pomeron is described in terms of a gluon ladder. The
cross section is proportional to $[\alpha_s(\bar q^2)]^2 \cdot [\bar x
g(\bar x, \bar q^2)]^2$, where $\alpha_s(\bar q^2)$ is the strong coupling
constant, $\bar x g(\bar x, \bar q^2)$ is the gluon momentum density in
the proton, $\bar x$ is the fraction of the proton's momentum carried by
the gluon ladder and $2 \bar q^2$, in the leading logarithm approximation,
is the upper limit for the virtuality of the two $t$-channel gluons of the
gluon ladder. Once $\alpha_s$ and the gluon momentum density are fixed,
the $W$ and $t$ dependences are determined.  The process under study is
sensitive to values of $\bar x \approx M_{\rho}^2/W^2 \approx 10^{-4}$ and
$\bar q^2 \approx 0.25~M_{\rho}^2 \approx 0.15$~GeV$^2$. The latter is
below the expected region of validity of the calculation; a
parametrisation for the product $[\alpha_s(\bar q^2)]^2 \cdot [\bar x
g(\bar x, \bar q^2)]^2$ was however found for which the model describes
all measured distributions well~\cite{tesi_luciano}. For this choice of
the parameters the cross section has a very weak dependence on $W$. The
two-pion invariant mass $M_{\pi\pi}$ was generated so as to reproduce,
after reconstruction, the measured distribution.

The second generator, LEVEC, was developed within the HERWIG
framework~\cite{herwig}. It assumes expression~(\ref{sigmat}) for the
$Q^2$ dependence of the cross section; the contribution of longitudinal
photons is neglected. The generated events were weighted such that all
other distributions (i.e. those over $W$, $M_{\pi\pi}$, $p_T^2$ and the
angular distributions of the decay pions), after detector simulation and
event reconstruction, have the same shape as those of the data.

For both programs, the angular distribution of the decay pions was assumed
to be that expected on the basis of SCHC~\cite{shilling-wolf}.

The acceptance for elastic $\rho^0$ production was calculated using both
the DIPSI and the LEVEC generators. Fig.~\ref{tacc}a, b and c respectively
show the acceptance as a function of $M_{\pi\pi}$, $W$ and $p_T^2$. The
acceptance includes the effects of the geometric acceptance of the
apparatus, of the detector efficiency and resolution, and of the trigger
and reconstruction efficiencies.  The trigger efficiency is $\approx
43\%$. The average acceptance is about 7\%.  The acceptance increases with
increasing $\pi \pi$ mass, has a broad maximum for $W~\ltap~70$~GeV and is
almost independent of the $\rho^0$ transverse momentum squared. In order
to convert the measured $dN/dp_T^2$ distribution to the
 differential cross section $d\sigma/dt$, the $p_T^2$ acceptance
(Fig.~\ref{tacc}c) was multiplied by a correction factor
$F$, which is, bin by bin, the
ratio of the $p_T^2$ and $t$ distributions
at the generator level. Fig.~\ref{tacc}d shows $F$ as a function of
$p_T^2$.

To produce the acceptance corrected cross sections, the generator DIPSI
was used. The difference between the results for the acceptance obtained
with the two generators was taken as an estimate of the systematic error
due to the model dependence of the acceptance calculation.

{}From the comparison of the distributions of the reconstructed and
generated events, the resolutions in $M_{\pi\pi}$ and $p_T^2$ were found
to be $\sigma_{M_{\pi\pi}} \simeq 20$ MeV~--~consistent with the mass
widths found for neutral strange particles reconstructed using the CTD
information~\cite{k0}~--~and $\sigma_{p_T^2} \simeq 0.01$~GeV$^2$.

The DIPSI and LEVEC simulations were also used to produce samples of
elastic $\omega$ and $\phi$ photoproduction events for the study of the
contamination from such processes.

The reaction $ep \rightarrow e \rho^0 X$, where $X$ is the hadronic state
resulting from the diffractive dissociation of the proton (see section
\ref{anal}), was simulated using PYTHIA~\cite{pythia}. A cross section of
the form $d^2\sigma /dt dM_X^2 \propto e^{bt}/M_X^\beta$, with $b=4.5$
GeV$^{-2}$, was assumed; the maximum value of $M_X$ was fixed by the
condition $M_X^2/W^2 \le 0.1$~\cite{chapin}. The exponent $\beta$ was
varied between 2 and 3, consistent with the result $\beta=2.20~\pm~0.03$
recently obtained at Fermilab~\cite{CDF} for the diffractive dissociation
of the proton in $\bar p p$ collisions. A second generator (RHODI) was
also used, based on the model calculation of ref.~\cite{jeff_misha}. In
both generators the $\rho^0$ decay angular distributions were assumed to
be the same as those of the elastic events.

\section{Backgrounds}
\label{anal}

After applying the selection criteria described in section \ref{cuts}, the
data still contain contributions from various background processes to the
reaction $ep \rightarrow e \pi^+ \pi^- p$:

\begin{itemize}
\item
inelastic $\rho^0$ production, $ep \rightarrow e  \rho^0 X$,
in which the proton diffractively dissociates into a hadronic
state $X$ not detected in the calorimeter;

\item
elastic production of $\omega$ and $\phi$ mesons;

\item
beam-gas interactions.
\end{itemize}

In order to estimate the contamination from the inelastic channel
$e p \rightarrow e \rho^0 X$, the energy distribution in the forward
calorimeter ($E_{FCAL}$) was studied. Fig.~\ref{fcale} shows the
distributions for both elastic and inelastic events as obtained from
Monte Carlo simulations based on the generators described above, along with the
distribution for the data. To obtain these plots, the cut
$E_{FCAL}<1$~GeV was not applied. The plot for elastic Monte Carlo events
(Fig. \ref{fcale}a) shows the FCAL energy spectrum resulting from noise
(mainly the uranium radioactivity).
Nearly all events have $E_{FCAL}<1$~GeV. For the
inelastic Monte Carlo sample as well as for the data, the $E_{FCAL}$ spectrum
extends to much higher values. Energy deposits in the forward calorimeter
greater than~1~GeV were therefore ascribed to
inelastic reactions in which part of the diffractively produced hadronic
state $X$ was detected in FCAL. The number of residual inelastic events
in the data with FCAL energy smaller than 1 GeV was then estimated as
\begin{eqnarray}
        N_{in} = \left\{
                   \frac{N(E_{FCAL}<1~\mbox{GeV})}{N(E_{FCAL}>1~\mbox{GeV})}
                 \right\}_{MC}
                 \times \left\{N(E_{FCAL}>1~\mbox{GeV})\right\}_{DATA}.
\nonumber
\end{eqnarray}
\noindent
The number of Monte Carlo and data events,
$\{N(E_{FCAL}>1~\mbox{GeV})\}_{MC}$ and
$\{N(E_{FCAL} > $ $1 ~\mbox{GeV})\}_{DATA}$, were computed in
the region
$1<E_{FCAL}<8$~GeV, where the inelastic Monte Carlo describes the data well.
The overall contamination integrated up to $|t| = 0.5$~GeV$^2$ was
estimated to be $11\% \pm 1\%~(\mbox{stat.}) \pm 6\%~(\mbox{syst.})$.
This was obtained using PYTHIA with $\beta=2.5$.
The systematic error reflects the sensitivity of the result to
the value of the exponent $\beta$, which was varied between 2 and 3, and to
the use of RHODI instead of PYTHIA.
As a check, the sum of the elastic and inelastic Monte Carlo distributions
of Fig.~\ref{fcale}a and~\ref{fcale}b were fitted to the data
of Fig.~\ref{fcale}c; the normalisations of the simulated distributions were
free parameters of the fit. The $\chi^2$ of the fit has a broad minimum
around $\beta=2.5$.  This method gave a $12\%$ contamination,
consistent with the result given above.
The result depends very little on the shape of the generated
$t$ distribution: the estimate of the contamination varies by less than
the quoted statistical error for a change of the
exponential slope $b$ between 4 and 6~GeV$^{-2}$.

The contamination was also studied as a function of $t$ and $W$. It was
found to vary from 6\% for $|t|<0.05$ GeV$^2$ to 19\% for $|t| \simeq 0.5$
GeV$^2$; it increases by $3 \pm 2\%$ as $W$ increases from 65 to 75 GeV.

The contamination due to the elastic production of $\phi$ and
$\omega$ mesons was estimated from Monte Carlo simulations.
The selection cuts described in section \ref{cuts}
were applied to the simulated events after reconstruction.
As an example, the contamination due to $\omega$ production
was estimated as
 $$\frac{A_{\omega}\sigma_{\omega}}
 {A_{\omega}\sigma_{\omega}+A_{\rho}\sigma_{\rho} },$$
where $A_{\omega}$ is the acceptance for elastic  $\omega$ events.
Assuming a cross section ratio of
$\sigma_{\omega}/\sigma_{\rho}=0.1$~\cite{totdat},
a contamination of $(1.3\pm0.2)\%$ is obtained.
A similar procedure applied to $\phi$ events results in an estimated
contamination of $(0.3 \pm 0.1)\%$, mainly due to the $\phi
\rightarrow 3 \pi$ decay mode. These contributions
were not subtracted but were included in the systematic error.
The contributions from inelastic
$\omega$ and $\phi$ production were negligible.

Electron beam-gas and proton beam-gas contaminations were deduced from the
pilot bunch event samples to which the cuts described in section
\ref{cuts} were applied. The number of events passing the cuts was then
scaled by the ratio between the electron (proton) current in the paired
bunches and the current in the electron (proton) pilot bunches. The
contamination due to electron-gas interactions was estimated to be 2.3
$\pm$ 0.5\%, while that due to proton-gas events was found to be 0.3 $\pm$
0.2\%.

All subsequent results are shown after subtraction of the contributions
from inelastic proton diffraction and beam-gas interactions.

\section{Results}

\subsection{Differential cross section $d\sigma/dM_{\pi\pi}$}
\label{mspec}

\noindent
Fig.~\ref{mplot} shows the acceptance corrected differential cross section
$d\sigma/dM_{\pi\pi}$.
The mass distribution is skewed compared to a Breit-Wigner distribution:
there is an enhancement of the low mass side and a suppression of the high
mass side. This distribution can
be understood in terms of the interference between the resonant
$\pi^+ \pi^-$ production and a non-resonant Drell-type background~\cite{drell}
as discussed by S\"oding \cite{soeding}.
In order to extract the contribution of
the resonant part of the differential cross section $d\sigma/dM_{\pi\pi}$,
we followed a procedure similar to that described in
refs.~\cite{bulos,park,omega}.
The function
\begin{eqnarray}
d\sigma/dM_{\pi\pi} = f_{\rho} \cdot BW_{\rho}(M_{\pi\pi})
+ f_I \cdot I(M_{\pi\pi}) + f_{PS}
\label{masf}
\end{eqnarray}
was fitted to the measured mass distribution. The term
\begin{eqnarray}
BW_{\rho}(M_{\pi\pi}) = \frac{M_{\pi\pi} M_{\rho}
\Gamma_{\rho}(M_{\pi\pi})} {(M_{\pi\pi}^2-M_{\rho}^2)^2 +
M_{\rho}^2 \Gamma_{\rho}^2(M_{\pi\pi})}
\label{breit}
\end{eqnarray}

\noindent
is a relativistic Breit-Wigner function, with a momentum dependent
width~\cite{jackson}
\begin{eqnarray}
\Gamma_{\rho}(M_{\pi\pi}) = \Gamma_0 \left(\frac{p^*}{p^*_0}\right)^3
\frac{M_{\rho}}{M_{\pi\pi}},
\label{gamma1}
\end{eqnarray}

\noindent
where $\Gamma_0$ is the width of the $\rho^0$, $p^*$ is the $\pi$
momentum in the $\pi \pi$ rest frame and $p^*_0$ is the value
of $p^*$ at the $\rho^0$ nominal mass $M_{\rho}$.
The function
\begin{eqnarray}
I(M_{\pi\pi}) = \frac{M_{\rho}^2-M_{\pi\pi}^2}{(M_{\pi\pi}^2-M_{\rho}^2)^2 +
M_{\rho}^2 \Gamma_{\rho}^2(M_{\pi\pi})}
\label{interf}
\end{eqnarray}

\noindent
is a parametrisation of the interference term.
The background term $f_{PS}$
was taken to be constant. The free parameters in the fit were $M_{\rho}$,
$\Gamma_0$ and the coefficients $f_{\rho}$, $f_I$ and $f_{PS}$.

The results of the fit are presented in table \ref{mtablfit} and in
Fig.~\ref{mplot}. The $\chi^2/ndf$ is 1.4, for $ndf$=13. The fitted values
of the $\rho^0$ mass and width are in good agreement with the accepted
ones~\cite{pdb}. The background term $f_{PS}$ is consistent with zero,
within a large error; similar results for $f_{PS}$ were obtained by
earlier experiments~\cite{bulos,park} using the functional
form~(\ref{masf}).

The contribution of the resonant term increases with $|t|$, ranging from
$86\%$ of the events for $|t|=0.01$~GeV$^2$ to $95\%$ for $|t|=0.5$
GeV$^2$.

The interference and the background terms were also studied as a function
of $W$ and of the decay pions' angular variables, $\cos{\theta_h}$ and
$\phi_h$, defined in section~\ref{angular}.  No dependence on these
variables was found.

The fit was repeated using different assumptions for the functional form
of $d\sigma/d M_{\pi\pi}$.

\begin{itemize}

\item Parametrisation~(\ref{masf}) is only an effective one, as it leaves
the interference term independent of the resonant and non-resonant terms,
which is strictly speaking inconsistent with the S\"oding mechanism.
We therefore fitted the following functional form to the invariant mass
distribution:

\begin{eqnarray}
d\sigma/dM_{\pi\pi} = \left|
A \frac{ \sqrt{ M_{\pi\pi} M_{\rho} \Gamma_{\rho}}}
{M_{\pi\pi}^2 - M_{\rho}^2 +i M_{\rho}\Gamma_{\rho}} + B \right|^2,
\label{wolf}
\end{eqnarray}

\noindent
where $A$, $B$, $M_{\rho}$ and $\Gamma_0$ were free parameters of the
fit. For $\Gamma_{\rho}$ expression~(\ref{gamma1}) was used.
The non-resonant amplitude $B$ was taken to be constant and real; it was also
constrained to be non-negative.
The results for the parameters are given in table~\ref{tabwolf}. The
$\chi^2/ndf$ was 1.4, with $ndf=14$.

\item The following alternative expressions for the width of the $\rho^0$
were adopted in the functions~(\ref{breit},\ref{interf}):
\begin{eqnarray}
\Gamma_{\rho}(M_{\pi\pi}) = \Gamma_0 \left(\frac{p^*}{p^*_0}\right)^3,
\label{gamma2}
\end{eqnarray}
\begin{eqnarray}
\Gamma_{\rho}(M_{\pi\pi}) = \Gamma_0 \left(\frac{p^*}{p^*_0}\right)^3
\frac{2} {1+(p^*/p^*_0)^2}.
\label{gamma3}
\end{eqnarray}

\item The Breit-Wigner was parametrised, following refs.~\cite{ballam,jackson},
as

\begin{eqnarray}
BW_{\rho}(M_{\pi\pi}) = \frac{1}{p^*}\frac{M_{\pi\pi} M_{\rho}
\Gamma_{\rho}(M_{\pi\pi})} {(M_{\pi\pi}^2-M_{\rho}^2)^2 +
M_{\rho}^2 \Gamma_{\rho}^2(M_{\pi\pi})}
\label{breit_ballam}
\end{eqnarray}

\noindent
and expression~(\ref{gamma3}) was used for the width.

\item The parametrisation given in ref.~\cite{egloff} was used:
\begin{eqnarray}
d\sigma/dM_{\pi\pi} = f_{\rho} \cdot BW_{\rho}(M_{\pi\pi}) \cdot
\left\{1+ C_1 \left[ (M_{\rho}/M_{\pi\pi})^2 -1 \right] + C_2
\left[ (M_{\rho}/M_{\pi\pi})^2 -1\right]^2\right\}.
\label{egloff1}
\end{eqnarray}
\noindent
The fit was repeated for the three mass dependent widths
(\ref{gamma1}),~(\ref{gamma2}) and~(\ref{gamma3}).

\item The phenomenological parametrisation proposed by Ross and
Stodolsky~\cite{stodolsky} was used:
\begin{eqnarray}
d\sigma/dM_{\pi\pi} = f_{\rho} \cdot BW_{\rho}(M_{\pi\pi}) \cdot
(M_{\rho}/M_{\pi\pi})^n + f_{PS},
\label{stodolsky}
\end{eqnarray}

\noindent
where the factor $(M_{\rho}/M_{\pi\pi})^n$ accounts for the skewing
of the shape of the $\rho^0$ signal. The term $f_{PS}$ was taken to be
constant. Here again the fit was repeated
for the three mass dependent widths
(\ref{gamma1}),~(\ref{gamma2}) and~(\ref{gamma3}).  The parameter $n$
was found to be
$n=4.9 \pm 0.5$, $n=5.8 \pm 0.5$ and $n=4.9 \pm 0.5$ for
the three forms of the width, respectively.

\end{itemize}

\noindent
In none of these cases did the quality of the fit change appreciably, as
can be seen from table~\ref{tabcross}, in which the values of $\chi^2/ndf$
obtained for the various functional forms are summarised.
The fitted values of the $\rho^0$ mass and width varied from
763 to 772 MeV and from 141 to 155 MeV, respectively.
As we discuss in the next section,
the values of the resonant part of the cross section were quite stable.

\subsection{Integrated $\gamma p \rightarrow \rho^0 p$ cross section}
\label{integrated}

The cross section $\sigma_{\gamma p \rightarrow \pi^+ \pi^- p}$ at $Q^2=0$,
integrated over the $M_{\pi\pi}$ and $t$
regions specified in~(\ref{kin}) and averaged over the range
$60<W<80$~GeV, can be obtained
from the data as $$ \sigma_{\gamma p \rightarrow \pi^+ \pi^- p} =
    \frac{ N_{\pi^+\pi^-}}{L \epsilon \Phi},$$
where
$N_{\pi^+ \pi^-}$ is the number of observed events after
background subtraction,
$\epsilon$ is the overall acceptance, $L$ is the effective luminosity
and $\Phi=0.02419$ is the effective photon flux factor (see eq. \ref{crs})
integrated
over the specified $W$ and $Q^2$ ranges.
In order to extract the cross section for the {\em resonant} process
$\gamma p \rightarrow \rho^0 p$,
it was assumed that the $\rho^0$ meson decays to $\pi^+ \pi^-$
with a $100\%$ branching ratio and the fit procedure described in
section~\ref{mspec} (with expressions~(\ref{masf}-\ref{interf}))
was used.
The resonant part of the total $\pi^+ \pi^-$ signal is given by the
parameter $f_{\rho}$ multiplied by the integral of the relativistic
Breit-Wigner curve, that is the area under the dotted curve in
Fig. \ref{mplot}. There is some arbitrariness in the choice of the
integration limits of the Breit-Wigner curve. The integral  was
carried out in the range
$2 M_{\pi}< M_{\pi\pi} < M_{\rho}+5\Gamma_0$, where the $\rho^0$ mass
and width values were taken from the fit; the quantity $M_{\pi}$ is the pion
mass. This requires an extrapolation beyond the measured
region.
The upper limit for the integration range approximately corresponds
to the mass of the nearest resonance, the $\rho (1450)$, with the same
quantum numbers and quark content as the $\rho^0$.
The value of the cross section,
for $2 M_{\pi}< M_{\pi\pi} < M_{\rho}+5\Gamma_0$,
$|t|<0.5$~GeV$^2$ and averaged over the region $60<W<80$~GeV,
was measured to be:
\begin{eqnarray}
\sigma_{\gamma p \rightarrow \rho^0 p} = 14.7\pm 0.4~(\mbox{stat.})
\pm2.4~(\mbox{syst.})~\mu\mbox{b}.
\label{result}
\end{eqnarray}
\noindent
The systematic error is dominated by the
uncertainties on the acceptance determination ($13\%$),
on the number of $\rho^0$ signal events (7\%),
and on the inelastic background determination ($7\%$).
If the integration is limited to the measured region,
$0.55 < M_{\pi\pi} < 1 $~GeV, the cross section value is 12.4 $\mu$b,
i.e. lower by a factor $\xi=1.19$.
If the integral is computed up to $M_{\rho}+4 \Gamma_0$, the cross section
decreases by $3\%$; if instead it is extended to
$M_{\rho}+6 \Gamma_0$, the cross section increases by $2\%$.

The uncertainty on the
acceptance determination has three main contributions:
\begin{enumerate}
\item the uncertainty on the calorimeter trigger efficiency near the
threshold ($ 9 \%$);
\item the difference between the results obtained with DIPSI and with LEVEC
($8 \%$);
\item the sensitivity of the results to the cuts, notably that on the minimum
number of CTD superlayers traversed by each track ($ 6\%$).
\end{enumerate}

The various alternative functional forms for $d\sigma/dM_{\pi\pi}$
described in the previous section were also used to extract the resonant part
of the signal. The values obtained
were centred around the result given in~(\ref{result}) but spanned the range
$\sigma_{\gamma p \rightarrow \rho^0 p}=13.6$-15.4~$\mu$b,
corresponding to a $^{+5}_{-8}\%$ maximum variation. The corresponding
variation of $\xi$ is $^{+8}_{-6}\%$.
The method of Spital and Yennie~\cite{spital} was
also used to obtain the cross section, yielding
\begin{eqnarray}
\sigma_{\gamma p \rightarrow \rho^0 p}=\frac{\pi \Gamma_0}{2}
\left.\frac{d\sigma}{ dM_{\pi\pi}}\right|_{M_{\rho}}=15.5 \pm 0.4~\mu\mbox{b},
\label{spital}
\end{eqnarray}

\noindent
where the $\rho^0$ mass and width were those given in table~\ref{mtablfit}.
The result obtained with the Spital and Yennie method depends
linearly on the value used for $\Gamma_0$; it is also sensitive to the
value of $M_{\rho}$, since $d\sigma / dM_{\pi\pi}$ is a steep
function of $M_{\pi\pi}$ in the region $M_{\pi\pi} \gtap 750$~MeV, as
can be seen from Fig.~\ref{mplot}.
If the values of $M_{\rho}$ and $\Gamma_0$ given in ref.~\cite{pdb} are used,
the
cross section changes by less than $1 \%$.
If $\Gamma_0$ is kept fixed
and $M_{\rho}$ is varied between 760 and 780~MeV, the corresponding change
in the cross section is $23 \%$.

Table~\ref{tabcross} summarises the results.
We have taken the spread into account by including a $\pm~7\%$ contribution
to the systematic uncertainty of the cross section.

The effect of real photon radiation by the incoming or the outgoing electron
and that of vacuum polarisation loops is to lower the measured value of
the cross section. The size of the correction
was estimated to be smaller
than 4\%~\cite{kurek}. The correction was not applied; instead a $4\%$
contribution was added to the systematic uncertainty.

Table~\ref{syserr} summarises the contributions to the systematic error.
The total systematic error was obtained by summing all contributions
in quadrature.

Result~(\ref{result}) for the cross section $\sigma_{\gamma p \rightarrow
\rho^0 p}$ is presented in Fig. \ref{xplot}a together with a partial
compilation of low energy measurements. The figure only includes results
explicitly corrected for the interference term and the non-resonant
background. We also show the ZEUS result~\cite{maciek}, obtained
indirectly using tagged photoproduction data from the 1992 data-taking
period. Fig.~\ref{xplot}b shows the result~(\ref{spital}) obtained with
the method of Spital and Yennie, along with a compilation of low energy
results obtained with the same technique. The dashed curve in both figures
is a parametrisation by Schuler and Sj\"{o}strand~\cite{schuler}; the $W$
dependence of the curve is based on Regge theory and on the assumption
that the intercept of the pomeron is $\alpha(0)=1+0.0808$ (the soft
pomeron). The same general trend of the data is seen in the two figures.
There are however differences in the results obtained by individual
experiments; these differences at least in part reflect the ambiguity in
the definition of the $\rho^0$ production cross section due to the finite
width of the $\rho^0$. The comparison between different experiments -- and
between the experimental results and the theoretical expectations --
should thus be taken with caution. The curve satisfactorily reproduces the
energy dependence of the data.

\subsection{Differential cross section $d\sigma/dt$}
\label{t}

\noindent
Fig.~\ref{tplot}a shows the acceptance corrected differential cross
section $d\sigma/dt$, integrated over the $\rho^0$ mass region
$2 M_{\pi} < M_{\pi\pi} < M_{\rho}+5\Gamma_0$.
It was obtained by multiplying the differential cross section for
the measured range $0.55<M_{\pi\pi} < 1$~GeV by the factor
$\xi=1.19$ discussed in section~\ref{integrated}. This assumes that,
in each $t$ bin, the ratio of the integral of the relativistic
Breit-Wigner distribution over the range
$2 M_{\pi} <M_{\pi\pi} < M_{\rho}+5\Gamma_0$ to that over the range
$0.55<M_{\pi\pi} < 1$~GeV is the same, i.e. that the mass and
the width of the $\rho^0$ are the same in each bin.
The contamination from inelastic $\rho^0$ production
was subtracted bin by bin. Background and interference terms
were also subtracted; their contribution was found by repeating the
fitting procedure described in section~\ref{mspec} in each $t$ bin,
fixing the values of the $\rho^0$ mass and width to those
given in table~\ref{mtablfit}.
Fig.~\ref{tplot}b shows
the results obtained for $d\sigma/dt$ by applying
the Spital and Yennie method in each $t$ bin.

The data were fitted with the function

\begin{eqnarray}
\frac{ d\sigma}{d|t|}   = A_t     \cdot e^{-b_t |t|}
\label{single}
\end{eqnarray}

\noindent
in the range
$|t| < 0.15$ GeV$^2$  and with the function

\begin{eqnarray}
\frac{ d\sigma}{d|t|} = A^{\prime}_{t}\cdot e^{-b^{\prime}_{t}
|t| + c^{\prime}_{t} t^2}
\label{double}
\end{eqnarray}

\noindent
in the range $ |t|<0.5$~GeV$^2$. Both functions describe the data well. The
results of the fits using expression~(\ref{double}) are shown
in Fig.~\ref{tplot}.
Table~\ref{fits} gives the values of the parameters obtained in the fits.
The difference between the results of the fits to the points of
Fig.~\ref{tplot}a and b
was taken as an estimate of the systematic uncertainty on the determination
of the resonant fraction of the cross section in each bin. The other
contributions to the systematic errors on $A_t$,
$A_t^{\prime}$  and $b_t$, $b_t^{\prime}$
are the uncertainties on the acceptance and on the
inelastic background determination.
All contributions to the systematic errors were summed in quadrature.
If the first bin is excluded from the fit
to the points of Fig.~\protect\ref{tplot}a, the values of the slopes
$b_t$ and $b_t^{\prime}$ increase by 9\% and 7\%, respectively;
the variation
is smaller for the fit of Fig.~\protect\ref{tplot}b.

Fig.~\ref{bplot} shows the result for the slope $b_t^{\prime}$ as obtained
from fitting eq.~(\ref{double}) to the points of Fig.~\ref{tplot}a
together with a compilation of fixed target results. All the results
shown were obtained  from fits of the
form~(\ref{double}); the data were explicitly corrected for interference and
non-resonant background, with the exception of those of ref.~\cite{jones}.
The data of refs.~\cite{jones,gladding} have the
somewhat large minimum $|t|$ values of 0.2 and 0.15~GeV$^2$, respectively.
For $W>4$~GeV, the results do not depend strongly on $W$,
as expected from Regge theory, which predicts a
logarithmic dependence of the slope on $W$ if one trajectory dominates.

\subsection{Decay angular distributions}
\label{angular}

The $\rho^0$ decay angular distributions can be used to determine elements
of the $\rho^0$ spin-density matrix~\cite{shilling-wolf}.

In the $s$-channel helicity frame, in which the $\rho^0$ direction in the
photon-proton centre-of-mass frame is taken as the quantisation axis, the
decay angular distribution $H(\cos\theta_h,\phi_h,\Phi_h)$ is a function
of the polar angle $\theta_h$ of the $\pi^+$ in the $\rho^0$
centre-of-mass frame, of the azimuthal angle $\phi_h$ between the decay
plane and the $\gamma$-$\rho^0$ plane (the $\rho^0$ production plane) and
of the angle $\Phi_h$ between the $\rho^0$ production plane and the
electron scattering one. For $t=t_{\min}$, the photon and the $\rho^0$ are
collinear and $\phi_h$ is not defined.

In the present experiment, the lepton scattering plane is not measured
since neither the recoil proton nor the scattered electron are detected.
Furthermore, the azimuthal angle $\phi_h$ can be determined only if the
direction of the virtual photon is approximated by that of the incoming
electron. It has been verified by Monte Carlo calculations that this is a
good approximation.
The experimental resolution in $\phi_h$ is approximately $40^{\circ}$; it
is a function of $t$ and improves with increasing $|t|$.
The resolution in
$\cos{\theta_h}$ is $\approx 0.03$. In the following we present the results for
the one dimensional distributions $H_{\cos \theta_h} (\cos \theta_h)$ and
$H_{\phi_h}(\phi_h)$, obtained from
$H(\cos\theta_h,\phi_h,\Phi_h)$  after integrating over $\phi_h$,
$\Phi_h$ and
over $\cos \theta_h$, $\Phi_h$, respectively. For
unpolarised or transversely polarised electrons
and a $J^P=1^{-}$ state decaying into two pions,
the functions $H_{\cos \theta_h} (\cos \theta_h)$ and $H_{\phi_h}(\phi_h)$ can
be
written as~\cite{shilling-wolf,joos}:
\begin{eqnarray}
H_{\cos \theta_h}=\frac{3}{4}[1-r_{00}^{04}+(3r_{00}^{04}-1)\cos^2{\theta_h}],
\label{W1}
\end{eqnarray}
\begin{eqnarray}
H_{\phi_h}=\frac{1}{2\pi}(1-2r_{1-1}^{04}\cos2\phi_h),
\label{W2}
\end{eqnarray}

\noindent
where $r_{00}^{04}$ and $r_{1-1}^{04}$ are $\rho^0$ density matrix elements.
In particular $r_{00}^{04}$ gives the probability that the $\rho^0$ is
longitudinally polarised. Assuming $s$-channel helicity
conservation (SCHC), $r_{00}^{04}$ can be related to $R$, the ratio
of the $\rho^0$ production cross sections for longitudinally and
transversely polarised virtual photons~\cite{joos}:
\begin{eqnarray}
R=\frac{1}{\varepsilon} \frac{r_{00}^{04}} {1-r_{00}^{04} },
\label{R}
\end{eqnarray}

\noindent
where $\varepsilon$ is the virtual photon polarisation, i.e. the
ratio of the longitudinally to transversely polarised photon fluxes.
The present data have  $\varepsilon = 0.998$, essentially constant over the
$W$ range covered by the measurement.

Fig.~\ref{cost} shows the differential cross sections $d\sigma/d\cos \theta_h$
and $d\sigma/d\phi_h$. The curves are the results of the fits of the
functions~(\ref{W1}) and~(\ref{W2}) to the data. A dominant $\sin^2\theta_h$
contribution is visible, indicating that the
$\rho^0$ mesons are mostly transversely polarised.
This is reflected by the result of the
fit, which yields $r_{00}^{04}=0.055 \pm 0.028$,
where the error is statistical only.
The fitted value of $r_{1-1}^{04}$ is $ 0.008 \pm 0.014$, consistent with
zero, as expected on the basis of SCHC.
If one uses expression~(\ref{R}) to determine  $R$ from the fitted value
of $r_{00}^{04}$, one obtains $R=0.06 \pm 0.03$. The average value of
$Q^2$ for the data discussed in this paper, computed
assuming the $Q^2$ dependence given in~(\ref{crs}),
is $\approx 0.1 M_{\rho}^2=
0.05$~GeV$^2$. This gives, using expression~(\ref{sigmal}),
$R=0.1$, consistent with our result.

\subsection{Total $\rho^0 p$  cross section}

By using the vector dominance model and the optical theorem, the measured
value of $d\sigma/dt$ at $t=0$ can be used to obtain the $\rho^0 p$ total
cross section.

For $\rho^0 p$ scattering the optical theorem reads:
\begin{eqnarray}
\left.\frac{ d\sigma_{\rho^0 p \rightarrow \rho^0 p}}{dt}\right|_{t=0}=
\frac{1+\eta ^2}{16\pi} \sigma_{tot}^2(\rho^0 p),
\label{optical}
\end{eqnarray}

\noindent
where $\eta$ is the ratio of the real to the imaginary part of the forward
$\rho^0 p$ scattering amplitude. We assume that $\eta=0$.
On the other hand, within VDM, elastic $\rho^0$ photoproduction is related
to the elastic $\rho^0 p$ cross section; in particular, for $t=0$:
\begin{eqnarray}
\left. \frac{d\sigma_{\gamma p \rightarrow \rho^0 p}}{dt}\right|_{t=0}=
\frac{4 \pi \alpha}{f_{\rho}^2}
\left. \frac{d\sigma_{\rho^0 p \rightarrow \rho^0 p}}{dt}\right |_{t=0},
\label{vmd1}
\end{eqnarray}

\noindent
where $4 \pi \alpha/f_{\rho}^2$ is the probability for the
$\gamma \rightarrow \rho^0$ transition.
We take $f_{\rho}^2/4\pi=2.20$~(cf.~e.g.~\cite{bible}, p.~393).

Using the intercept $d\sigma_{\gamma p \rightarrow \rho^0 p}/dt|_{t=0}=
133 \pm 11 \pm 27$~$\mu$b/GeV$^2$ given in section~\ref{t}, and combining
equations~(\ref{vmd1}) and~(\ref{optical}), one finds $\sigma_{tot}(\rho^0
p)=28.0 \pm 1.2~(\mbox{stat.}) \pm 2.8~(\mbox{syst.})~\mbox{mb}$, where
the errors reflect only the uncertainty on the measured value of the
intercept. The systematic error does not include the uncertainties from
the model dependence of the assumptions made nor from the values of $\eta$
and $4\pi/f_{\rho}^2$.

The result is consistent with those found at lower energies (see
e.g.~\cite{bible,nmc_quasi}). It is also in agreement with the expected
value of $27.8$~mb at $W=70$~GeV, obtained from the
parametrisation~\cite{schuler} of $\sigma_{\rho^0 p}^{tot} \approx 1/2
[\sigma^{tot}(\pi^+ p) + \sigma^{tot}(\pi^- p)]=13.63 (W^2)^{0.0808} +
31.79 (W^2)^{- 0.4525}$, based on the soft pomeron, the additive quark
model~\cite{aqm} and on fits~\cite{dl1} to $\pi p$ data at centre-of-mass
energies ranging between $6$ and 25~GeV.

\section{Summary}

The ZEUS detector at HERA has been used to study the photoproduction
process $\gamma p \rightarrow \rho^0 p$. The integrated cross section as
well as the differential cross sections $d\sigma/dM_{\pi\pi}$ and
$d\sigma/dt$ at an average photon-proton centre-of-mass energy $\langle W
\rangle =$ 70~GeV have been measured.

The $\pi \pi$ mass spectrum is skewed, compared to a relativistic
Breit-Wigner distribution, as also observed at low energy. The integrated
$\gamma p \rightarrow \rho^0 p$ cross section, for $60<W<80$~GeV,
$|t|<0.5$~GeV$^2$ and $2 M_{\pi}< M_{\pi\pi} < M_{\rho}+5\Gamma_0$, is
$14.7\pm 0.4~(\mbox{stat.}) \pm2.4~(\mbox{syst.})~\mu\mbox{b}$. This
result, in conjunction with the measurements at low energy, is consistent
with the weak energy dependence expected on the basis of Regge theory and
a pomeron intercept of $1.08$ (the soft pomeron). This is at variance with
the behaviour of the cross section for elastic photoproduction of $J/\psi$
mesons~\cite{zeus_psi} and elastic production of $\rho^0$ mesons for
$Q^2>7$~GeV$^2$~\cite{zeus_hiq2rho}.

The differential cross section $d\sigma/dt$ has an approximately
exponential shape with a slope consistent with the results obtained by low
energy experiments with $W>4$~GeV. This is also in accord with Regge
theory, which expects a logarithmic dependence of the slope on $W$.

The $\rho^0$ decay angular distributions have been studied.  The $\rho^0$
mesons are mainly produced in a transversely polarised state;  the
probability to find the $\rho^0$ with longitudinal polarisation is
$r_{00}^{04}=0.055\pm 0.028$. The value of $r_{1-1}^{04}$ is $0.008 \pm
0.014$, in accord with $s$-channel helicity conservation (SCHC). If SCHC
is assumed, one obtains $R=0.06 \pm 0.03$ for the ratio of the $\rho^0$
production cross sections for longitudinally and transversely polarised
virtual photons, consistent with the value expected if VDM is assumed.

{}From the measured value of $d\sigma/dt$ at $t=0$, within the VDM framework
and by using the optical theorem, a value $\sigma^{tot}_{\rho^0 p}=28.0
\pm 1.2~(\mbox{stat.}) \pm 2.8~(\mbox{syst.})~\mbox{mb}$ for the total
$\rho^0p$ cross section at $W=70$~GeV is found, in agreement with
extrapolations of the pion-proton total cross section data based on Regge
theory and a soft pomeron.

\section*{Acknowledgements}

We thank the DESY Directorate for their strong support and encouragement.
The remarkable achievements of the HERA machine group were essential for
the successful completion of this work, and are gratefully appreciated.
It is a pleasure to thank N.N.~Nikolaev, M.G.~Ryskin and A.~Sandacz for
many enlightening discussions. We are also very grateful to K.~Kurek for
his calculation of the radiative corrections.

\newpage
\begin{table}
\begin{center}
\begin{tabular}{lcc} \hline \hline
Parameter           & value            & stat. error       \\ \hline
$M_{\rho}$          & 0.764 GeV        & 0.003 GeV         \\
$\Gamma_0$          & 0.142 GeV        & 0.007 GeV         \\
$ f_{\rho}$         & 9.86 $\mu$b     & 0.23 $\mu$b       \\
$f_{I}$             & 3.47 $\mu$b GeV & 0.34 $\mu$b GeV  \\
$f_{PS}$            & 0.0  $\mu$b/GeV & 5.5  $\mu$b/GeV  \\ \hline \hline
\end{tabular}
\end{center}
\caption{Results of the fit to the mass spectrum
for $60~<W<80$ GeV and $|t|<0.5$ GeV$^2$
(expressions~(\protect\ref{masf}-\protect\ref{interf})).
Only statistical errors are given.}
\label{mtablfit}
\end{table}

\vspace{5cm}

\begin{table}
\vspace{5cm}
\begin{center}
\begin{tabular}{lcc} \hline \hline
Parameter           & value            & stat. error       \\ \hline
$M_{\rho}$          & 0.764 GeV        & 0.003 GeV         \\
$\Gamma_0$          & 0.148 GeV        & 0.007 GeV         \\
$A$            & $-3.10~\mu$b$^{1/2}$     & 0.04~$\mu$b$^{1/2}$      \\
$B$             & 2.0  $\mu$b$^{1/2}$ GeV$^{-1/2}$ & 0.2 $\mu$b$^{1/2}$
GeV$^{-1/2}$  \\ \hline \hline
\end{tabular}
\end{center}
\caption{Results of the fit to the mass spectrum
for $60~<W<80$ GeV and $|t|<0.5$ GeV$^2$ (expression~(\protect\ref{wolf})).
Only statistical errors are given.}
\label{tabwolf}
\end{table}

\begin{table}
\begin{center}
\begin{tabular}{lccc} \hline \hline
Functional form &$\chi^2/ndf~(ndf)$  &$\sigma_{\gamma p \rightarrow
\rho^0 p}$ [$\mu$b] & $\sigma_{\gamma p \rightarrow \rho^0 p}$ [$\mu$b] \\
                &
&$0.55<M_{\pi\pi}<1$~GeV                         &
$2M_{\pi}<M_{\pi\pi}<M_{\rho}+5\Gamma_0$  \\ \hline
Expressions~(\protect\ref{masf},\protect\ref{breit},\protect\ref{gamma1},\protect\ref{interf})        & 1.46 (13) & $12.4 \pm 0.3$ & $14.7 \pm 0.4$ \\
Expressions~(\protect\ref{masf},\protect\ref{breit},\protect\ref{gamma2},\protect\ref{interf})        & 1.64 (13) & $11.5 \pm 0.2$ & $14.5 \pm 0.3$ \\
Expressions~(\protect\ref{masf},\protect\ref{breit},\protect\ref{gamma3},\protect\ref{interf})        & 1.45 (13) & $12.3 \pm 0.6$ & $14.3 \pm 0.7$ \\
Expressions~(\protect\ref{masf},\protect\ref{breit_ballam},\protect\ref{gamma2},\protect\ref{interf}) & 1.46 (13) & $12.1 \pm 0.6$ & $13.6 \pm 0.6$ \\
Expressions~(\protect\ref{wolf},\protect\ref{gamma1})
                      & 1.38 (14) & $11.9 \pm 0.3$ & $14.3 \pm 0.4$ \\
Expressions~(\protect\ref{egloff1},\protect\ref{gamma1})
                      & 1.45 (13) & $12.4 \pm 0.5$ & $14.7 \pm 0.6$ \\
Expressions~(\protect\ref{egloff1},\protect\ref{gamma2})
                      & 1.49 (13) & $12.0 \pm 0.5$ & $15.4 \pm 0.6$ \\
Expressions~(\protect\ref{egloff1},\protect\ref{gamma3})
                      & 1.47 (13) & $12.3 \pm 0.5$ & $14.4 \pm 0.6$ \\
Expressions~(\protect\ref{stodolsky},\protect\ref{gamma1})
                      & 1.45 (13) & $12.3 \pm 0.6$ & $14.6 \pm 0.8$ \\
Expressions~(\protect\ref{stodolsky},\protect\ref{gamma2})
                      & 1.45 (13) & $11.9 \pm 0.6$ & $15.2 \pm 0.8$ \\
Expressions~(\protect\ref{stodolsky},\protect\ref{gamma3})
                      & 1.46 (13) & $12.3 \pm 0.6$ & $14.3 \pm 0.7$ \\
Expression~(\protect\ref{spital})
                      &   -       &  -             & $15.5 \pm 0.4$ \\ \hline
\hline
\end{tabular}
\end{center}
\caption{Values of $\chi^2/ndf$ and results for the cross section
$\gamma p \rightarrow \rho^0 p$
obtained by fitting the different functional forms described in the text
to the data of fig.~\protect\ref{mplot}. The indicated
errors are the statistical ones only.
}
\label{tabcross}
\end{table}

\begin{table}
\begin{center}
\begin{tabular}{lc}                            \hline \hline
 Contribution from                 &  Error \\ \hline
 Luminosity                        &  3\% \\
 Acceptance: trigger efficiency    &  9\%   \\
 Acceptance: model dependence      &  8\%   \\
 Acceptance: sensitivity to cuts   &  6\%   \\
 Inelastic background subtraction  &  7\%   \\
 Background due to elastic $\omega$ and $\phi$ production &  1\%   \\
 Procedure to extract the resonant part of the cross section & 7\% \\
 Radiative corrections             &  4\%   \\
 \hline
 Total                             & 17\%   \\
 \hline \hline
\end{tabular}
\end{center}
\caption{Individual contributions to the systematic error on the integrated
cross section.
}
\label{syserr}
\end{table}

\begin{table}
{\footnotesize
\begin{center}
\begin{tabular}{lll|lll} \hline \hline
           &  Data from Fig.~\protect\ref{tplot}a
&                       &           &
   Data from Fig.~\protect\ref{tplot}b &    \\
\hline
$A_t    $ & $= 139   \pm 6~   \mbox{(stat.)} \pm 26~   \mbox{(syst.)}$ &
$\mu$b/GeV$^2$       &
 $A_t    $ & $= 159   \pm 6~   \mbox{(stat.)} \pm 28~   \mbox{(syst.)}$ &
$\mu$b/GeV$^2$ \\
$b_t    $ & $=  10.4 \pm 0.6~ \mbox{(stat.)} \pm  1.1~ \mbox{(syst.)}$ &
GeV$^{-2}$           & $b_t    $ &
 $=  11.2 \pm 0.5~ \mbox{(stat.)} \pm  1.1~ \mbox{(syst.)}$ & GeV$^{-2}$ \\
\hline
$A^{\prime}_t    $ & $= 133 \pm 11~  \mbox{(stat.)} \pm 27 ~
\mbox{(syst.)}$ & $\mu$b/GeV$^2$ & $A^{\prime}_t    $
 & $= 155  \pm 14~  \mbox{(stat.)} \pm 29~ \mbox{(syst.)}$ &
$\mu$b/GeV$^2$ \\
$b^{\prime}_t    $ & $= 9.9 \pm 1.2~ \mbox{(stat.)} \pm 1.4~
\mbox{(syst.)}$ & GeV$^{-2}$     & $b^{\prime}_t    $
 & $= 11.1 \pm 1.5~ \mbox{(stat.)} \pm 1.4~ \mbox{(syst.)}$ & GeV$^{-2}$ \\
$c^{\prime}_t    $ & $= 2.3 \pm 2.8 ~\mbox{(stat.)} $                  &
GeV$^{-4}$     &
 $c^{\prime}_t    $ & $= 4.7  \pm 3.3 ~\mbox{(stat.)} $
& GeV$^{-4}$ \\
\hline \hline
\end{tabular}
\end{center}
\caption{The results of the fits to the data of Fig.~\protect\ref{tplot}a
(left) and Fig.~\protect\ref{tplot}b (right) with
function~(\protect\ref{double}).
For an explanation of the symbols see text.
}
\label{fits}
}
\vspace{12cm}
\end{table}
\newpage

\newpage

\begin{figure}[t]
\begin{center}
\leavevmode
\hbox{%
\epsfxsize = 17cm
\epsffile{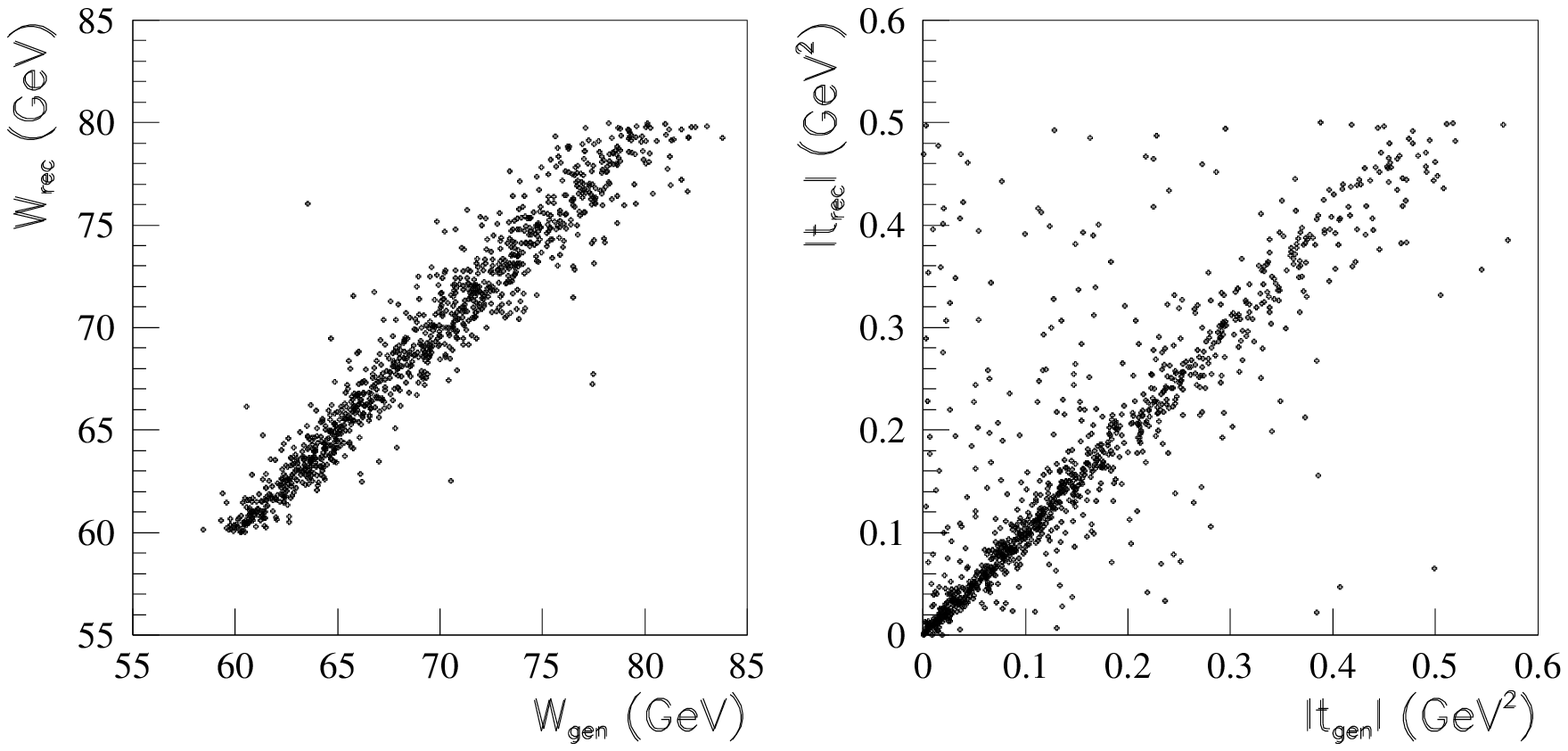}}
\end{center}
\vspace{-1.0cm}
\caption{Scatter plot of reconstructed versus generated values
of $W$ and $|t|$ for Monte Carlo events.
}
\label{papercomp}
\efig

\newpage
\begin{figure}[t]
\begin{center}
\leavevmode
\hbox{%
\epsfxsize = 14.8cm
\epsffile{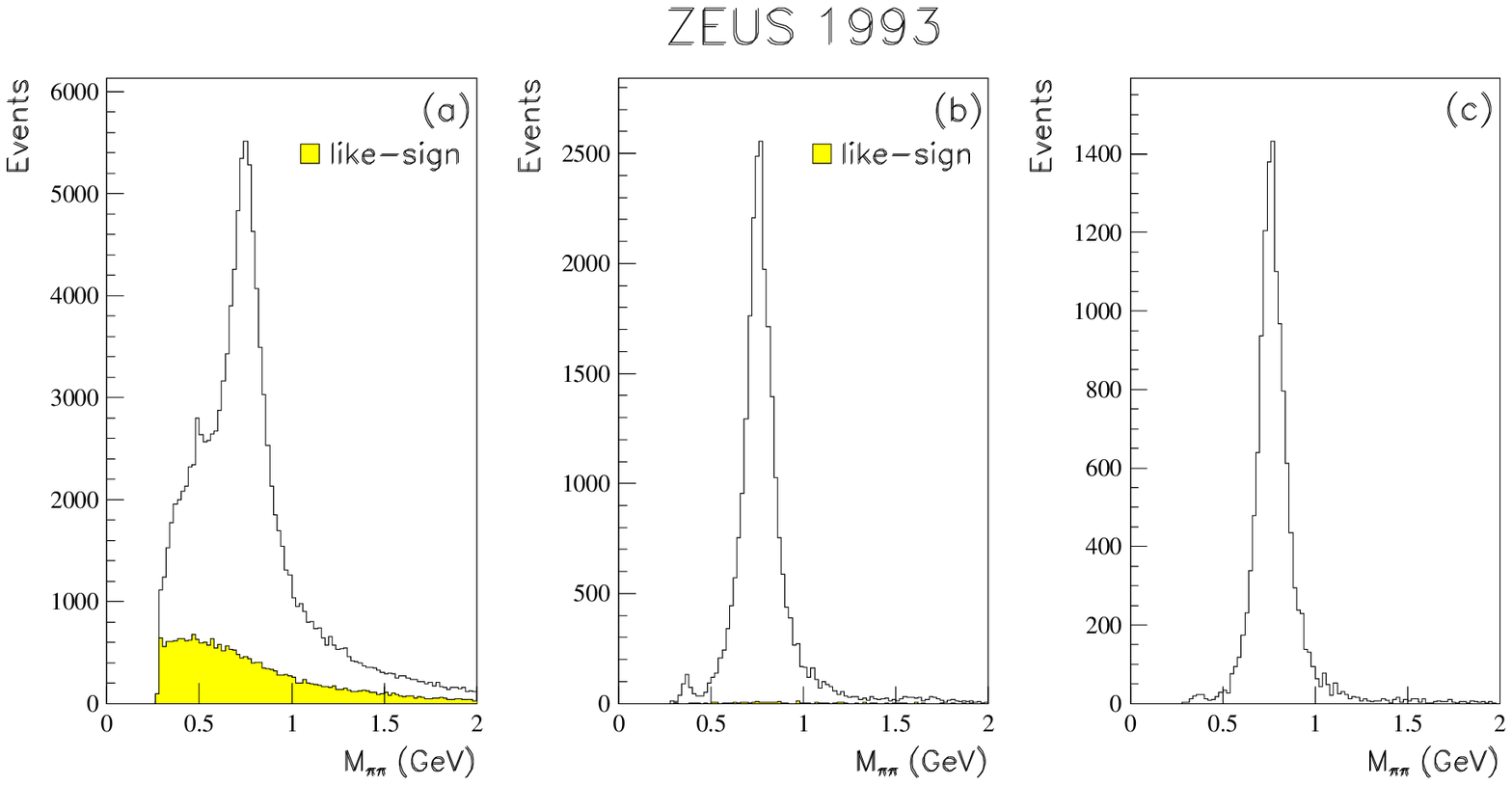}}
\end{center}
\vspace{-6.7cm}
\caption{(a) The invariant $\pi^+ \pi^-$ mass spectrum for all two
track events selected by the untagged photoproduction trigger. (b) The
distribution after applying all cuts, except those
on $W$, $M_{\pi\pi}$ and $p_T^2$.
The shaded area  in both plots (hardly visible in (b)) is the mass
distribution of the like-sign pion pairs.
(c) The distribution after all cuts, including
those on $W$ and $p_T^2$ (but not that on $M_{\pi\pi}$).
}
\label{massbef}
\efig

\newpage
\begin{figure}[t]
\begin{center}
\leavevmode
\hbox{%
\epsfxsize = 15cm
\epsffile{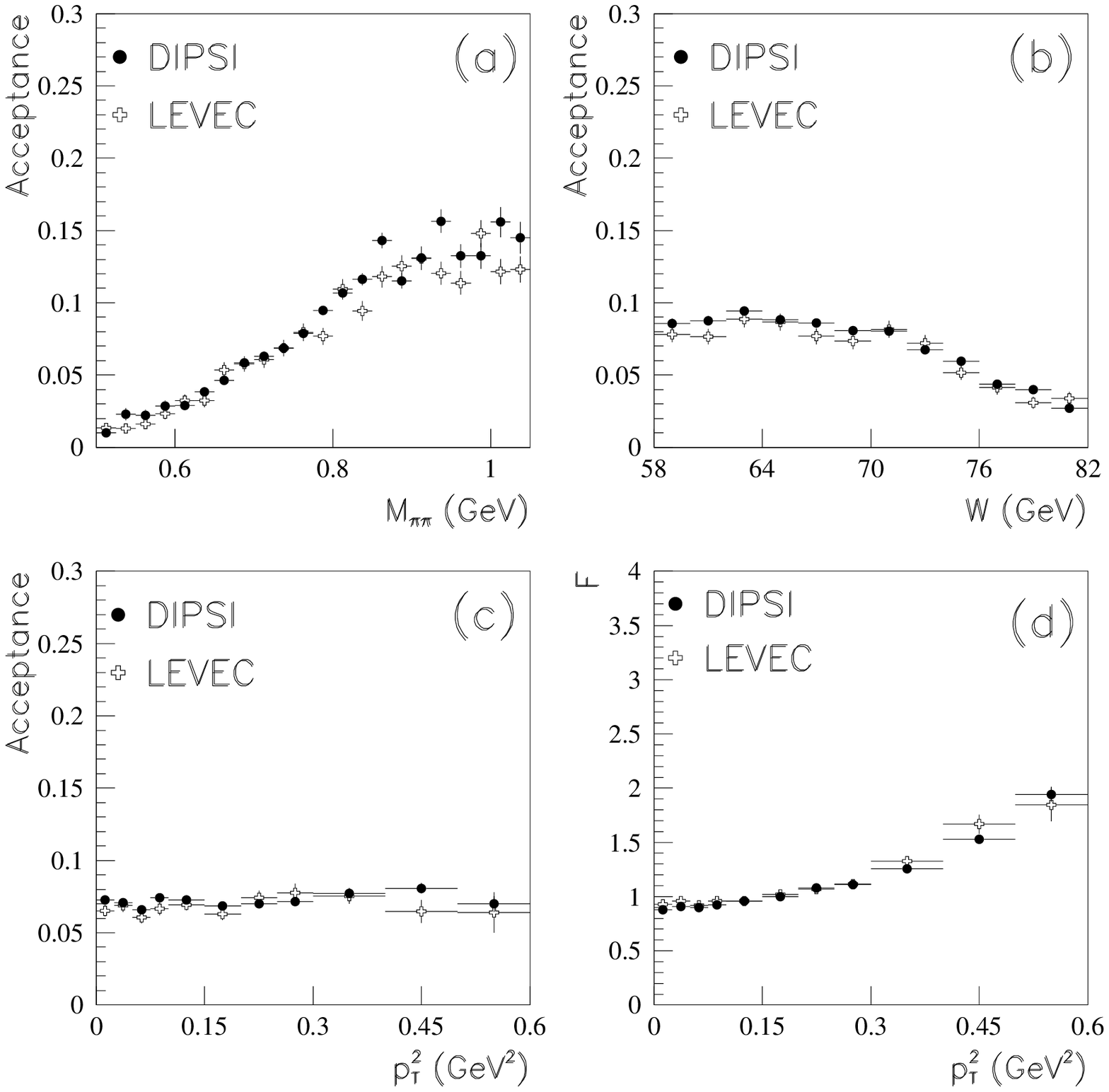}}
\end{center}
\vspace{-3.5cm}
\caption{Acceptance as a function of $M_{\pi\pi}$ (a), $W$
(b) and $p_T^2$ (c) obtained with the DIPSI and the LEVEC generators.
The quantity $F$ plotted in (d) is defined in the text.
Only statistical errors are shown. The horizontal bars indicate the size
of the bins.}
\label{tacc}
\efig

\newpage
\begin{figure}[t]
\begin{center}
\leavevmode
\hbox{%
\epsfxsize = 14.0cm
\epsffile{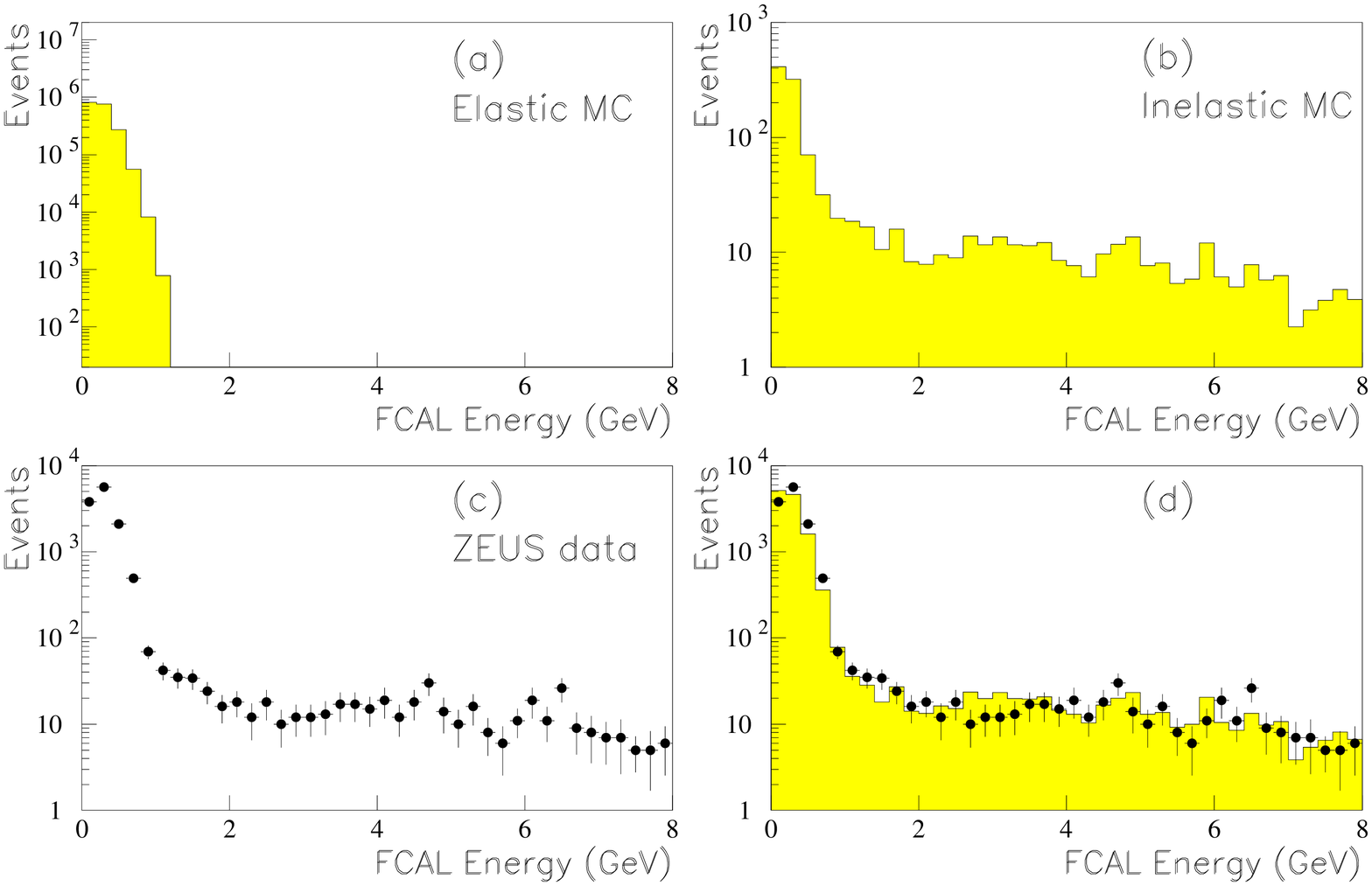}}
\end{center}
\vspace{-0.5cm}
\caption{The energy spectrum in the forward calorimeter for:
(a) elastic production ($ep \rightarrow e \rho^0 p$) simulated with DIPSI;
(b) inelastic production ($ep \rightarrow e \rho^0 X$) simulated with
PYTHIA; (c) data; (d) a mixture of 89\% elastic and 11\% inelastic
Monte Carlo events (histogram) compared with data (dots).
}
\label{fcale}
\efig

\newpage
\begin{figure}[t]
\begin{center}
\leavevmode
\hbox{%
\epsfxsize = 17cm
\epsffile{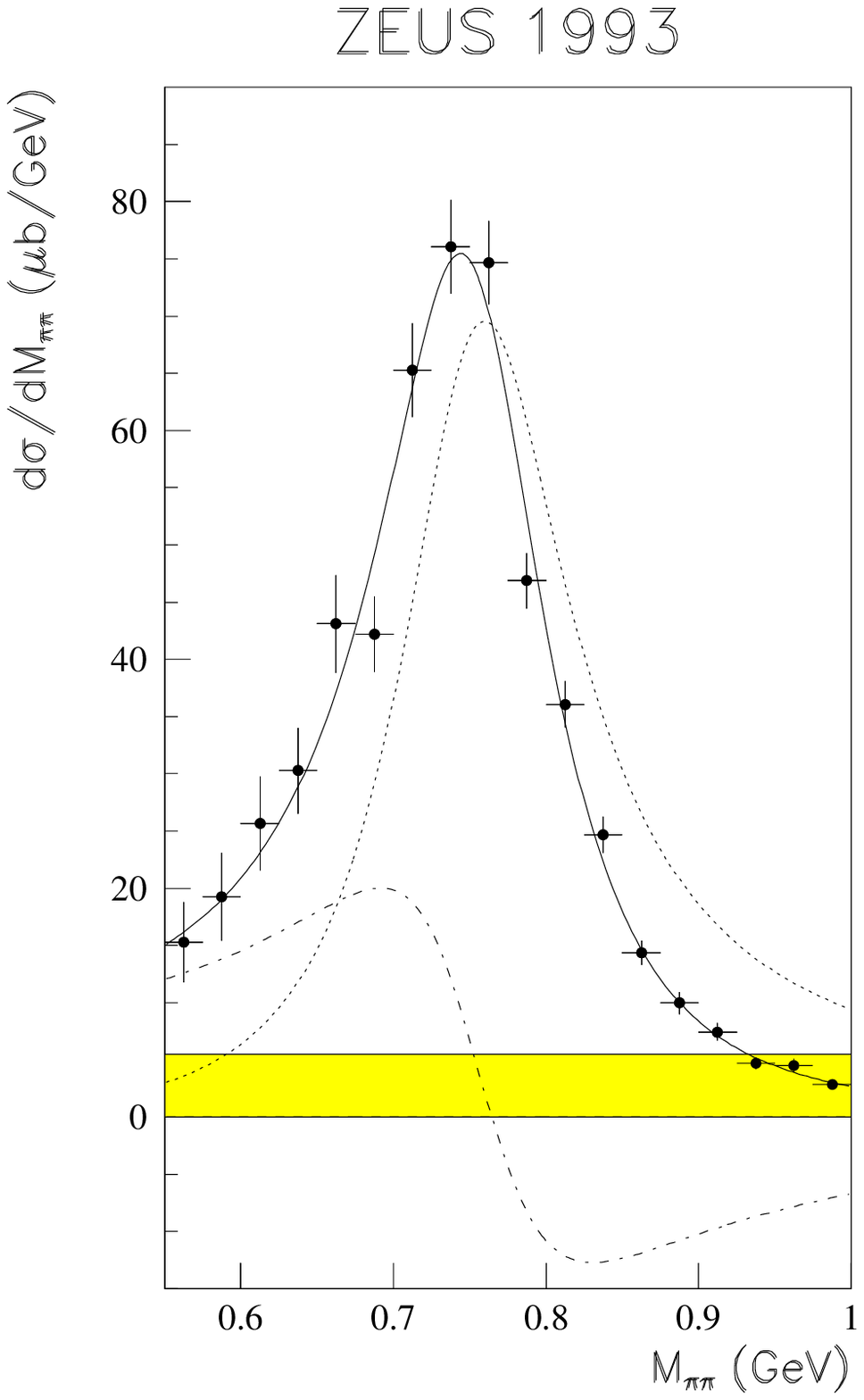}}
\end{center}
\vspace{-4cm}
\caption{The acceptance corrected differential cross section
$d\sigma/dM_{\pi\pi}$
for $60<W<80$~GeV and $|t|<0.5$~GeV$^2$. The points represent
the ZEUS data and the curves indicate the results of the fit to the data
using
expressions~(\protect\ref{masf}-\protect\ref{interf}).
The dotted line shows the contribution of the
Breit-Wigner term and
the dash-dotted line  that of the interference term;
the shaded band indicates the
size of the statistical uncertainty on the background term $f_{PS}$.
The solid curve is the sum of these three terms. Only statistical errors
are shown. The horizontal bars indicate the size of the bins.}
\label{mplot}
\efig

\newpage
\begin{figure}[t]
\begin{center}
\leavevmode
\hbox{%
\epsfxsize = 15cm
\epsffile{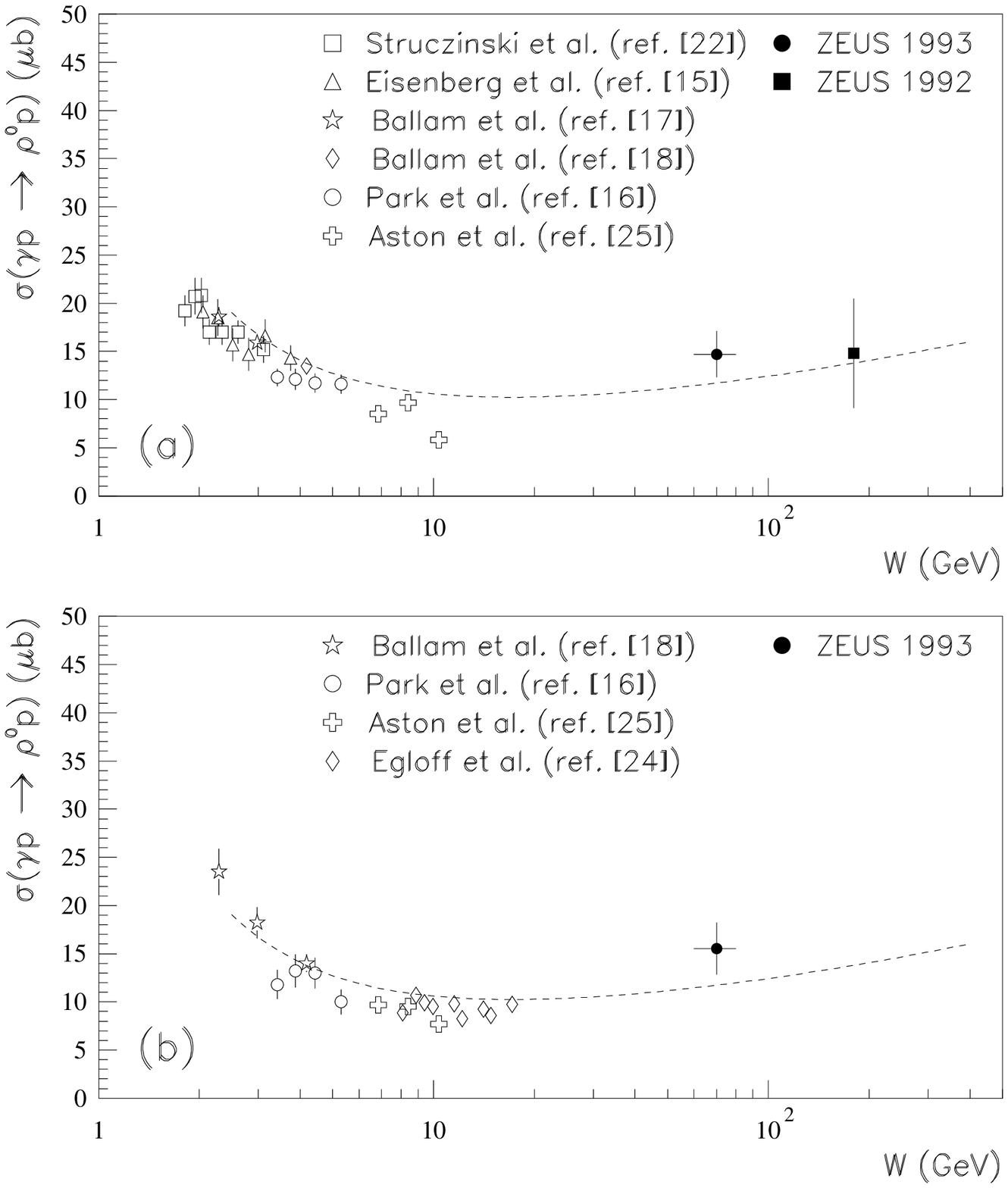}}
\end{center}
\vspace{-1.5cm}
\caption{(a) The integrated cross section $\sigma_{\gamma p \rightarrow
\rho^0 p}$
as a function of the centre-of-mass energy $W$. The ZEUS result (labelled
``ZEUS 1993") is that given in~(\protect\ref{result}) for the range
$2M_{\pi}<M_{\pi\pi}<M_{\rho}+5\Gamma_0$, $|t|<0.5$~GeV$^2$. The ZEUS
result~\protect\cite{maciek} is also shown.
The dashed line is the  parametrisation from~\protect\cite{schuler}. The
vertical
error bars
of the ZEUS points indicate the quadratic sum of statistical and systematic
errors. The horizontal bar indicates the size of the $W$ region covered
by the
present measurement. (b) Same as (a) except that the results  shown were
obtained with the Spital and Yennie method.
}
\label{xplot}
\efig

\newpage
\begin{figure}[t]
\begin{center}
\leavevmode
\hbox{%
\epsfxsize = 17cm
\epsffile{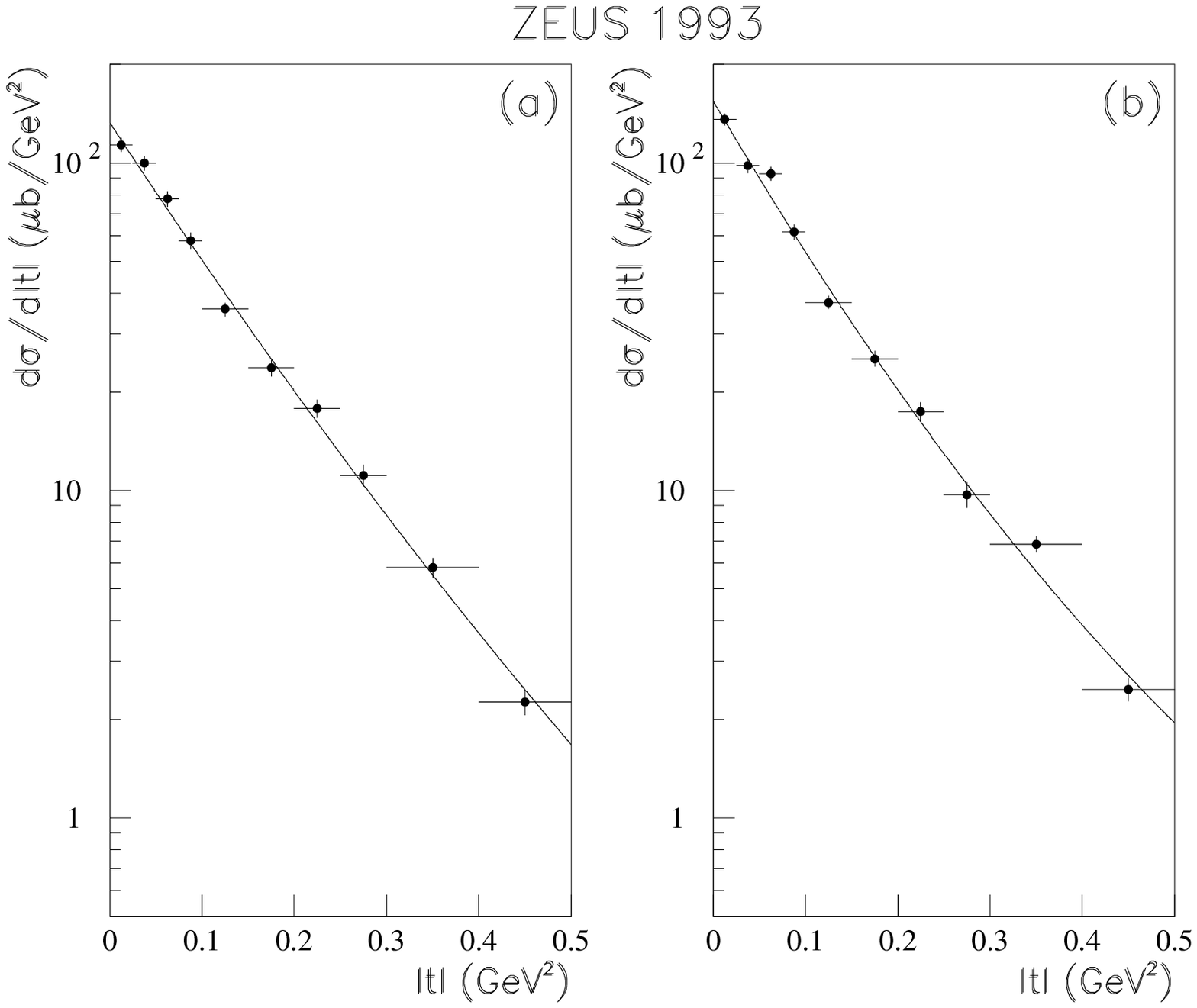}}
\end{center}
\vspace{-5.5cm}
\caption{(a) The differential cross section
$d\sigma/dt$ for $\gamma p \rightarrow \rho^0 p$ in the mass
range $2M_{\pi}<M_{\pi\pi}<M_{\rho}+5\Gamma_0$ and $60 <W<80$~GeV. (b)
The differential cross section
$d\sigma/dt$ for $\gamma p \rightarrow \rho^0 p$
for $60 <W<80$~GeV, as obtained by applying the Spital and Yennie
method in each $t$ bin.
The continuous lines represent the results of the
fit with the functional form~(\protect\ref{double}).
Only statistical errors are shown. The horizontal bars indicate
the size of the bins.}
\label{tplot}
\efig

\newpage
\begin{figure}[t]
\begin{center}
\leavevmode
\hbox{%
\epsfxsize = 15cm
\epsffile{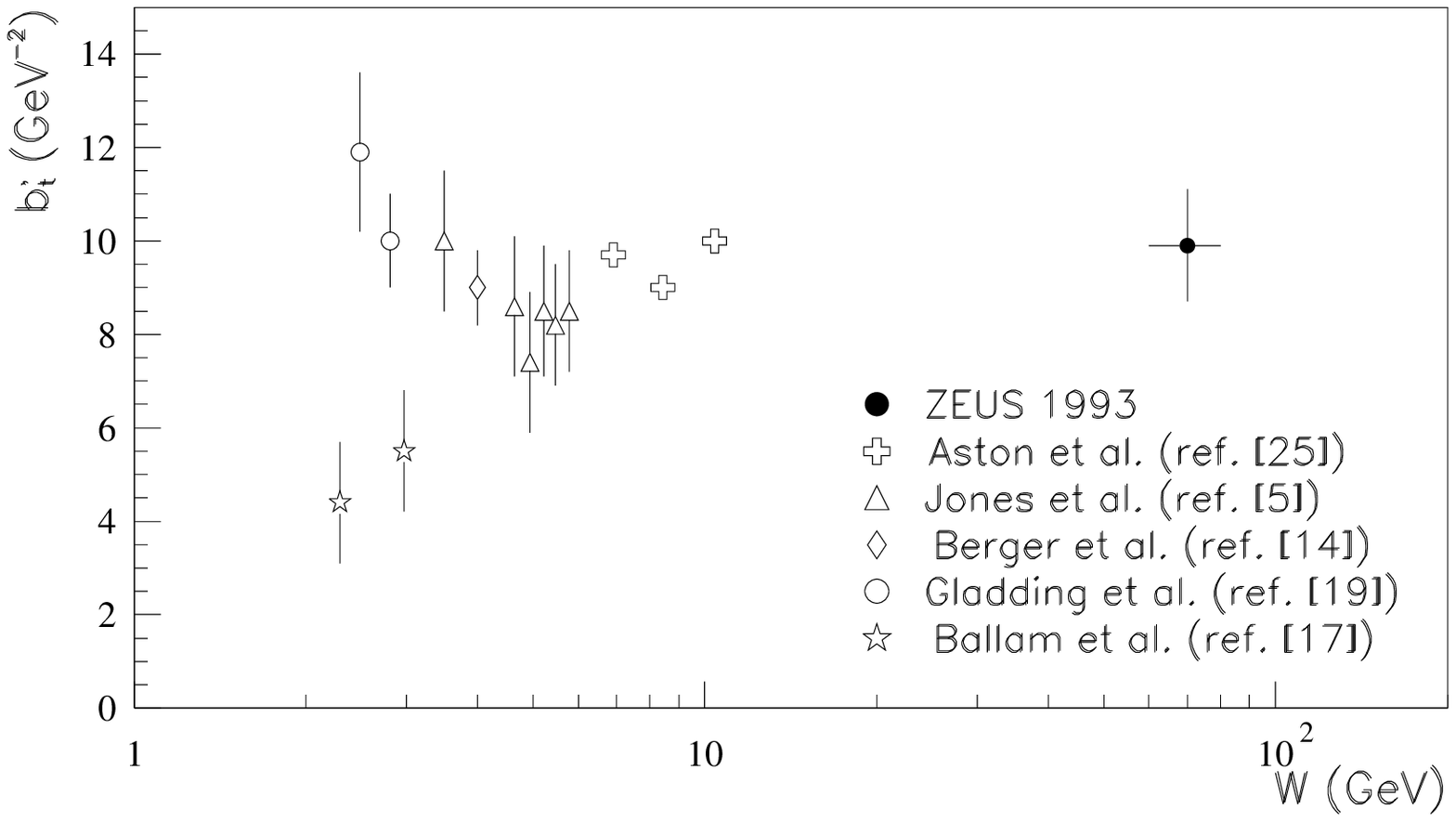}}
\end{center}
\vspace{-1.0cm}
\caption{The slope $b_t^{\prime}$ as a function of $W$ as obtained from
fits with
function~(\protect\ref{double}). The ZEUS result is that found by fitting
to the data of Fig.~\protect\ref{tplot}a.
The vertical error bars indicate statistical errors only.
The horizontal bar indicates the size of $W$ region covered by the
present measurement.
}
\label{bplot}
\efig

\newpage

\begin{figure}[t]
\begin{center}
\leavevmode
\hbox{%
\epsfxsize = 15cm
\epsffile{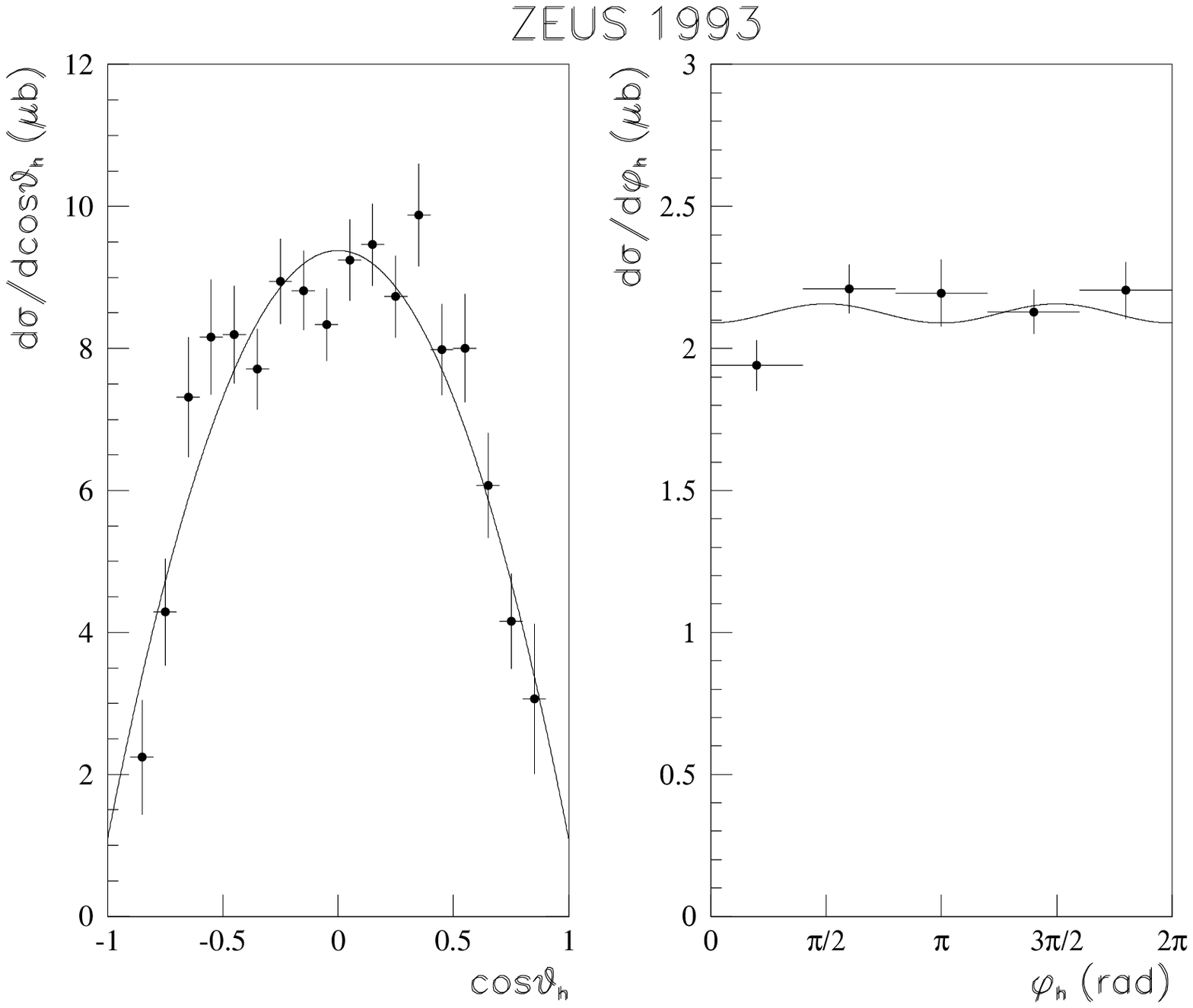}}
\end{center}
\vspace{-5.0cm}
\caption{The differential cross sections $d\sigma/d\cos\theta_h$ and
$d\sigma/d\phi_h$. The continuous lines represent the results of the
fits discussed in the text. Only statistical errors are shown.
The horizontal bars indicate the size of the bins.
}
\label{cost}
\efig


\begin{thebibliography}{99}
\addcontentsline{toc}{chapter}{Bibliografy}

\bibitem{bible} For a review, see e.g.\\
T.H. Bauer et al., Rev. Mod. Phys. {\bf 50} (1978) 261.

\bibitem{blechscmidt} H. Blechschmidt et al., Nuovo Cimento {\bf 52A} (1967)
1348.
\bibitem{lanzerotti} L.J. Lanzerotti et al., Phys. Rev. {\bf 166} (1968) 1365.
\bibitem{erbe} ABBHHM Collab., R. Erbe et al., Phys. Rev. {\bf 175} (1968)
1669.
\bibitem{jones} W.G. Jones et al., Phys. Rev. Lett. {\bf 21} (1968) 586.
\bibitem{bulos} F. Bulos et al., Phys. Rev. Lett. {\bf 22} (1969) 490.

\bibitem{mcclellan} G. McClellan et al., Phys. Rev. Lett. {\bf 22} (1969) 374.
\bibitem{bingham} SBT Collab., H.H. Bingham et al., Phys. Rev. Lett. {\bf 24}
(1970) 955.
\bibitem{ballam00} SBT Collab., J. Ballam et al.,  Phys. Rev. Lett. {\bf 24}
(1970) 960.
\bibitem{alvensleben} DESY-MIT Collab., H. Alvensleben et al., Nucl. Phys.
{\bf B 18} (1970) 333.
\bibitem{anderson} R. Anderson et al., Phys. Rev. {\bf D1} (1970) 27.
\bibitem{davier} M. Davier et al., Phys. Rev. {\bf D1} (1970) 790.
\bibitem{mcclellan1} G. McClellan et al., Phys. Rev. {\bf D4} (1971) 2683.
\bibitem{berger} C. Berger et al., Phys. Lett. {\bf 39B} (1972) 659.
\bibitem{eisenberg} SWT Collab., Y. Eisenberg et al., Phys. Rev. {\bf D5}
(1972) 15.
\bibitem{park} J. Park et al., Nucl. Phys. {\bf B 36} (1972) 404.
\bibitem{ballam0} SBT Collab.,  J. Ballam et al., Phys. Rev. {\bf D5} (1972)
545.
\bibitem{ballam} SBT Collab, J. Ballam et al., Phys. Rev. {\bf D7} (1973) 3150.

\bibitem{gladding} G.E. Gladding et al., Phys. Rev. {\bf D8} (1973) 3721.

\bibitem{alexander}  G. Alexander et al., Nucl. Phys. {\bf B 69}  (1974) 445.
\bibitem{alexander1} G. Alexander et al., Nucl. Phys. {\bf B 104} (1976) 397.
\bibitem{struczinski} W. Struczinski et al., Nucl. Phys. {\bf B 108} (1976) 45.
\bibitem{barish} B. Barish et al., Phys. Rev. {\bf D9} (1974) 566.
\bibitem{egloff} R. M. Egloff et al, Phys. Rev. Lett. {\bf 43} (1979) 657.
\bibitem{omega} OMEGA Collab., D. Aston et al., Nucl. Phys. {\bf B 209} (1982)
56.

%Q2.ne.0
\bibitem{joos} P. Joos et al., Nucl. Phys. {\bf B 113} (1976) 53.
\bibitem{chio}  CHIO Collab., W. D. Shambroom et al., Phys. Rev. {\bf D26}
(1982) 1.
\bibitem{emc1}  EMC Collab., J.J. Aubert  et al., Phys. Lett. {\bf 161B} (1985)
203.
\bibitem{emc2}  EMC Collab., J. Ashman et al., Z. Phys. {\bf C 39} (1988) 169.

\bibitem{e665}
 E665 Collab., G. Y. Fang  et al., preprint FERMILAB-Conf. 93/305 (1993);\\
 G. Y. Fang, in ``Proceedings of the XIII International Conference on
 Particle and Nuclei", Perugia, Italy, 28th June-3rd July 1993,
 editor~A.~Pascolini, World Scientific, Singapore (1994)~p.~332.

\bibitem{nmc} NMC Collab., M. Arneodo et al., Nucl. Phys. {\bf B 429} (1994)
503.

\bibitem{maciek}
ZEUS Collab., M.Derrick et al., Z. Phys. {\bf C 63} (1994) 391.

\bibitem{saku}
J.J Sakurai, Phys. Rev. Lett. {\bf 22} (1969) 981.

\bibitem{goulianos} See e.g. \\
K. Goulianos, Phys. Rep. {\bf101} (1983) 169.

\bibitem{dl}
 A. Donnachie and P.V. Landshoff, Phys. Lett. {\bf B 185} (1987) 403;\\
 A. Donnachie and P.V. Landshoff, Nucl. Phys. {\bf B 311} (1989) 509;\\
 P.V. Landshoff, Nucl.Phys. B (Proc.Suppl.) {\bf 18C} (1990) 211;\\
 P.V. Landshoff, in ``Proc. of Joint LP Symposium and Europhysics Conference
  on HEP'', Geneva 1991, editors S. Hegarty, K. Potter and E. Quercigh,
  World Scientific, Singapore, 1992, vol.2, p.~363.

\bibitem{cudell}
 J.R. Cudell, Nucl. Phys. {\bf B 336} (1990) 1.

\bibitem{misha} M.G. Ryskin, Z. Phys. {\bf C 57} (1993) 89
and private communication.

\bibitem{boris}
 B.Z. Kopeliovich et al., Phys. Lett. {\bf B 324} (1994) 469.

\bibitem{brodsky}
 S.J. Brodsky et al., Phys. Rev. {\bf D50} (1994) 3134.

\bibitem{kolya} J. Nemchik, N.N. Nikolaev and B.G. Zakharov, Phys. Lett.
{\bf B 341} (1994) 228.

\bibitem{ginzburg} I.F. Ginzburg, D.Yu. Ivanov and V.G. Serbo,
``Semihard quasi diffractive production of neutral mesons
by off shell photons'', to appear in Nucl. Phys. B.

\bibitem{status93}
The ZEUS Detector, Status Report, DESY (1993).

\bibitem{VXD}
C. Alvisi et al., Nucl. Instrum. Methods {\bf A~305} (1991) 30.

\bibitem{CTD}
N.~Harnew et al., Nucl. Instrum. Methods {\bf A~279} (1989) 290;\\
C.B.~Brooks et al., Nucl. Instrum. Methods {\bf A~283} (1989) 477;\\
B.~Foster et al., Nucl. Instrum. Methods {\bf A~338} (1994) 254.

\bibitem{CAL}
M.~Derrick et al., Nucl. Instrum. Methods {\bf A~309} (1991) 77;\\
A.~Andresen et al., Nucl. Instrum. Methods {\bf A~309} (1991) 101;\\
A.~Bernstein et al., Nucl. Instrum. Methods {\bf A~336} (1993) 23.

\bibitem{lumi}
D. Kisielewska et al., ``Fast Luminosity Monitoring at HERA",
DESY-HERA report 85-25 (1985),\\
J.~Andruszk\'ow et al., DESY report DESY~92-066 (1992).

\bibitem{tesi}
M. Costa, ``Fotoproduzione di mesoni $\rho$ in interazioni $ep$ ad HERA'',
Tesi di Dottorato, University of Torino (1994), unpublished (in Italian).

\bibitem{dipsi} M. Arneodo, L. Lamberti and M. G. Ryskin, to be submitted to
Comp. Phys. Comm.
\bibitem{tesi_luciano}
L. Lamberti, ``Fotoproduzione esclusiva di mesoni vettori nell'esperimento
ZEUS ad HERA", Tesi di Dottorato, University of Torino (1995), unpublished (in
Italian).


\bibitem{herwig}
G. Marchesini et al., Comp. Phys. Comm. {\bf 67} (1992) 465;\\
B.R. Webber, in ``Proceedings of the Workshop `Physics at HERA",
DESY, 29-30 October 1991, editors~W. Buchm\"{u}ller and G. Ingelman, p.~1354;\\
L. Stanco, ibidem, p.~1363.

\bibitem{shilling-wolf} K. Schilling et al., Nucl. Phys. {\bf B 15} (1970)
397;\\
K. Schilling and G. Wolf, Nucl. Phys. {\bf B 61}
(1973) 381.

\bibitem{k0} ZEUS Collab., M.Derrick et al., DESY report DESY~95-084
(1995), submitted to Z. Phys. C.

\bibitem{pythia}
T. Sj\"{o}strand and M. Bengsston, Comp. Phys. Comm. {\bf 43} (1987) 367;\\
M. Bengsston and T. Sj\"{o}strand, Comp. Phys. Comm. {\bf 46} (1987) 43;\\
T. Sj\"{o}strand, in ``Proceedings of the Workshop Physics at HERA",
DESY, 29-30 October 1991, editors~W. Buchm\"{u}ller and G. Ingelman, p.~1405.

\bibitem{chapin}
T.J. Chapin et al., Phys.~Rev. {\bf D31} (1985) 17.


\bibitem{CDF} CDF Collab., F. Abe et al., Phys. Rev.
{\bf D50} (1994) 5535.

\bibitem{jeff_misha} J. Forshaw and M. Ryskin, DESY report DESY~94-162 (1994),
Rutherford Laboratory preprint RAL-94-058 (1994), and private
communication.

\bibitem{totdat}
``Total Cross-Sections for Reactions of High Energy Particles'',
Landolt-B\"ornstein, New Series, Vol. 12b, editor~H.~Schopper (1987).

\bibitem{drell} S.D. Drell, Phys. Rev. Lett. {\bf 5} (1960) 278.

\bibitem{soeding}
P. S\"{o}ding, Phys. Lett. {\bf 19} (1966) 702.

\bibitem{jackson}
J.D. Jackson, Nuovo Cimento {\bf 34} (1964) 1644.
\bibitem{pdb} Review of Particle Properties, Phys. Rev. {\bf D50} (1994) 1173.
\bibitem{stodolsky} R. Ross and V. Stodolsky, Phys. Rev. {\bf 149} (1966) 1172.
\bibitem{spital} R. Spital and D.R. Yennie, Phys. Rev. {\bf D9} (1974) 126.
\bibitem{kurek} K. Kurek, private communication.

\bibitem{schuler}
G.A.~Schuler and T.Sj\"ostrand,
Phys.~Lett. {\bf B 300} (1993) 169;\\
G.A.~Schuler and T.Sj\"ostrand, Nucl. Phys. {\bf B 407} (1993)~539.


\bibitem{nmc_quasi} M. Arneodo et al., Phys. Lett.  {\bf B 332} (1994) 195.

\bibitem{aqm}
E.M. Levin and L.L. Frankfurt, JETP Letters {\bf 2} (1965) 65;\\
H.J. Lipkin and F. Scheck, Phys. Rev. Lett. {\bf 16} (1966) 71;\\
J.J.J. Kokkedee, ``The Quark Model", W.A. Benjamin, New York (1969).

\bibitem{dl1} A. Donnachie and P.V. Landshoff, Phys. Lett. {\bf B 296}
(1992) 227.

\bibitem{zeus_psi} ZEUS Collab., M.Derrick et al.,
Phys. Lett. {\bf B 350} (1995) 120.

\bibitem{zeus_hiq2rho} ZEUS Collab., M.Derrick et al., DESY report DESY
95-133 (1995), submitted to Phys. Lett. B.


\end{thebibliography}
\end{document}